\documentclass[english]{iopart}
\usepackage[T1]{fontenc}
\usepackage[latin9]{inputenc}
\usepackage[dvipsnames]{xcolor}
\usepackage{amstext}
\usepackage{graphicx}
\usepackage{hyperref}
\usepackage{bm}
\usepackage{appendix}
\usepackage{ifthen}
\usepackage{float}
\usepackage{multirow}
\usepackage{booktabs}
\usepackage{lipsum} 
\usepackage[ruled,vlined]{algorithm2e}
\usepackage{cite}
\usepackage{todonotes}
\usepackage[switch]{lineno} 
\usepackage{soul}
\usepackage{cancel}

\makeatletter
\usepackage{iopams}
\usepackage{setstack}

\usepackage{textcomp}

\usepackage{iopams}
\expandafter\let\csname equation*\endcsname\relax
\expandafter\let\csname endequation*\endcsname\relax

\expandafter\let\csname substack\endcsname\relax
\expandafter\let\csname subarray\endcsname\relax
\expandafter\let\csname endsubarray\endcsname\relax
\expandafter\let\csname sideset\endcsname\relax
\expandafter\let\csname underset\endcsname\relax
\expandafter\let\csname overset\endcsname\relax
\expandafter\let\csname ddddot\endcsname\relax
\expandafter\let\csname dddot\endcsname\relax
\expandafter\let\csname boxed\endcsname\relax
\expandafter\let\csname leftroot\endcsname\relax
\expandafter\let\csname uproot\endcsname\relax

\usepackage{amsmath}

\makeatother

\usepackage{babel}
\begin{document}

\newcounter{printtext}
\setcounter{printtext}{0}
\newcounter{printfig}
\setcounter{printfig}{0}

\title{Model-based electron density profile estimation and control, applied to ITER}

\author{T.~O.~S.~J.~Bosman$^{1,2}$, M.~van~Berkel$^{1}$, and M.~R.~de~Baar$^{1,2}$}

\address{\noindent {\footnotesize{}$^{1}$DIFFER - Dutch Institute for Fundamental
Energy Research, PO Box 6336, 5600HH Eindhoven, The Netherlands}}

\address{\noindent {\footnotesize{}$^{2}$Eindhoven University of Technology,
Dept.~of Mechanical Engineering, Control Systems Technology Group,
PO Box 513, 5600MB Eindhoven, The Netherlands}}


\begin{abstract}
In contemporary magnetic confinement devices, the density distribution is sensed with interferometers and actuated with feedback controlled gas injection and open-loop pellet injection. This is at variance with the density control for ITER and DEMO, that will depend mainly on pellet injection as an actuator in feed-back control. 
This paper presents recent developments in state estimation and control of the electron density profile for ITER using relevant sensors and actuators.
As a first step, Thomson scattering is included in an existing dynamic state observer. 
Second, model predictive control is developed as a strategy to regulate the density profile while avoiding limits associated with the total density (Greenwald limit) or gradients in the density distribution (e.g. neo-classical impurity transport). Simulations show that high quality density profile estimation can be achieved with Thomson Scattering and that the controller is capable of regulating the distribution as desired.
\end{abstract}

\submitto{Journal of Physics Communications}

\maketitle
\ioptwocol

\section{Introduction}\label{sec:intro}
Future reactors, such as ITER and DEMO, aim to have a net energy gain \cite{Shimomura1999}. This requires these reactors to operate close to physical limits and to optimize quantities such as temperature, density, and confinement. In practice, shot-to-shot differences in reactor conditions will always be present and the optimal control action will vary subject to the plasma scenario and plasma state. Hence, active plasma feedback control is essential in achieving these requirements \cite{Pironti2005a}.

In a tokamak, the produced fusion power directly correlates to the density in the hot core of the reactor \cite{Wesson2011}, particle transport and non-inductive current drive depend on the gradient of the spatial particle distribution \cite{Angioni2009,Kessel1994}, a.k.a the (particle) density profile. Furthermore, turbulence and impurity transport depend on the logarithmic gradient of the density profile \cite{Angioni2014,Horton1999,Garbet2006} and the density is subject to limits that can lead to detrimental plasma instabilities when violated \cite{Greenwald2002,Boozer2012,Zohm2015}. Consequently, for optimal and reliable high-performance operation of fusion reactors, particle density profile estimation and control are mandatory \cite{Gribov2007,Loarte2007,Kurihara2008,Biel2015}.

To synthesize such controllers, model-based control design is required as experimental time for tuning and validation on reactors such as ITER will be scarce and expensive \cite{Humphreys2015,DeVries2018}. Model-based design has already successfully been applied to synthesize controllers for plasma shape \cite{Albanese2005} and profiles in a tokamak \cite{Boyer2014,Barton2015,Moreau2015,Maljaars2017}. Especially, in \cite{Maljaars2015}, a controller for the current density profile is presented that can handle plasma limits and actuator constraints simultaneously. The design procedure requires a control-oriented model  of the input-output dynamics, i.e., a model that captures the main dynamic relationships between inputs (actuators) and outputs (plasma quantities) to be controlled. 

Recently, the control-oriented model RAPDENS was developed for the particle transport dynamics in a tokamak \cite{Blanken2018}. This model has been used to perform density profile estimation in TCV and AUG with interferometry and spectroscopy \cite{Blanken2018,Blanken2019}, to design a controller for the volume-averaged density during the ITER ramp-up with pellets and gas \cite{Ravensbergen2018}, to estimate the core density in density control experiments with pellets in AUG \cite{Lang2018}, and to derive a controller for the volume-average density in TCV using gas injection \cite{Blanken2019}.

However, a controller for the particle density distribution,: a) that is capable of handling shot-to-shot differences; b) can handle state physical limits; c) and uses multiple 
actuators is one key research and development aspect still to be solved for a fusion reactor.

In this work, a two-step model-based approach is proposed to develop such a density profile controller for ITER. The first is the estimation (or reconstruction) of the particle density profile with an observer and relevant sensors. The second is the design and validation of a model-predictive controller for the density profile using two pellet injectors.

For the first step, RAPDENS \cite{Blanken2018} is applied to ITER and extended with a Thomson scattering (TS) output model. Subsequently, the model and synthetic TS measurements are used together in an extended version of the dynamic state observer (DSO) proposed in \cite{Blanken2018} to reconstruct the density profile. Simulations of the observer are used to determine the quality of the achieved profile estimation for an H-mode, 15 MA, 5.3 T, DT ITER baseline discharge. 

For the second step, model-predictive control (MPC) is applied to synthesize a controller for the density distribution in ITER using pellet injection as main actuator. For the first time, constraints on the density profile due to impurity transport are taken into consideration in the control problem. Simulations using the lightweight control-oriented plasma simulator RAPDENS are used to assess the effectiveness of the controller.

The remainder of this paper is structured as follows. Section \ref{sec:RAPDENS} briefly summarizes the control-oriented model used for design and simulation. The forward TS model and the DSO are presented in Section \ref{sec:DSO}. The density reconstruction results are presented in Section \ref{sec:EKF results}. The controller design methodology is described in Section \ref{sec:MPC method} and the results of control simulations are shown in Section \ref{sec:MPC results}. This paper is concluded by a discussion of the work in Section \ref{sec:discussion}.

\section{Control-oriented plasma simulator}\label{sec:RAPDENS}
In this work, we have used and extended the RAPDENS model (originally proposed in \cite{Blanken2018}) for the DSO and the synthesis of the MPC controller. A summary of the model is given here.\\

\begin{figure}[hbtp]
\centering
\includegraphics[scale=1.5]{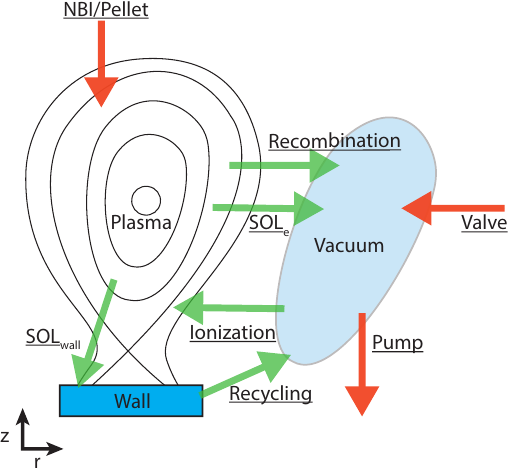}
\caption{Overview of the modeled processes in the RAPDENS model. The particles inside the tokamak are attributed either to the plasma, the wall or the vacuum. Green arrows represent the modeled particle fluxes between the inventories. Orange arrows represent particle fluxes crossing the system boundaries. Adapted from \cite{Bosman2021}}.
\label{fig:RAPDENS overview}
\end{figure}

The RAPDENS model is a lightweight, control-oriented model of the electron particle transport in a tokamak \cite{Blanken2018}. It is a multi-inventory model in which the particles inside the vessel are attributed to one of three inventories: the plasma, the wall, or the vacuum. Note that the vacuum denotes the region surrounding the plasma that is filled with neutrals. However, the name "vacuum" is kept for continuity and conherence with previous papers on RAPDENS. The model consists of a 1D-PDE for the evolution of the flux-surface averaged electron density 
and two OD-ODE's for the evolution of the wall and neutral vacuum inventories. Radial particle transport is modeled with a drift-diffusion model with ad-hoc chosen transport coefficients. The processes that drive the particle exchange between these inventories are modeled in a heuristic fashion. An overview is given in figure \ref{fig:RAPDENS overview}. Additionally, it contains representations of the fueling with neutral beam, pellet and gas injection. A summary of the model equations is given is \ref{app:RAPDENS}. The full details and derivations of the model equations are outside the scope of this paper and can be found in \cite{Blanken2018}.

For the applications reported in this paper, RAPDENS is used to formulate a system of nonlinear discrete-time ODEs that describes the evolution of the plasma density due to relevant actuator inputs. 
The internal variables of the model (states), the external parameters, the considered actuators inputs, and the system of ODEs are briefly discussed next.

\subsubsection{States:} In RAPDENS, the flux-surface averaged electron density distribution $n_e(\rho,t)$ is discretized in space using finite elements (see e.g. \cite{Hughes2000}), i.e, the electron density is approximated by 
\begin{equation}
n_e(\rho,t) = \sum_{\alpha = 1}^m\Lambda_{\alpha}(\rho)b_{\alpha}(t),
\label{eq:b-spline discretization}
\end{equation}
where $\rho$ is the normalized toroidal flux. The basis function $\Lambda_{\alpha}: [0,\rho_e]\rightarrow[0,1], \alpha = 1,2,\hdots,m$ are defined as cubic B-splines with finite support \cite{Boor1980} ($\rho_e > 1$ being the artificially prolonged flux label to encompass the scrape-off layer. This is discussed in more details in \cite{Blanken2018}). The variables $b_{\alpha}(t) \in \mathbb{R}^m$ are time dependent spline coefficients. 
These coefficients together with the time dependent wall and vacuum inventories, respectively denoted $N_w(t)$ and $N_v(t)$, comprise the state vector $x(t) \in \mathbb{R}^{n_x}$ with $n_x = m+2$, i.e., the state vector is given by 
\begin{equation}
x(t) = \begin{bmatrix}b(t)&N_w(t) & N_v(t) \end{bmatrix}^\top.
\label{eq:RAPDENS state vector}
\end{equation}

\subsubsection{External parameters:} 
RAPDENS requires external parameters to be provided in an external parameter vector $p(t) \in \mathcal{P}$ defined as
\begin{equation}
p = [c_D \;\;c_H\;\; T_{e,b}\;\; I_p\;\; V\;\;V'\;\;\kappa\;\;\psi_{axis}\;\;\psi_{LCFS}\;\;  G_1\;\; G_0],\label{eq:parameter vector}
\end{equation}
where $c_D$ indicates limited ($c_D=0$) or diverted ($c_D=1$) plasma, $c_H$ implies low confinement ($c_H=0$) or high confinement ($c_H=1$) regime, $T_{e,b}$ is the electron temperature at the last-closed flux surface (LCFS), $I_p$ is the plasma current, $V$ is the plasma volume, $V'=dV/dt$, and $\kappa$ is the plasma elongation, $\psi_{axis}$ and $\psi_{LCFS}$ are the equilibrium $\psi(R,Z)$ evaluated at the magnetic axis and LCFS, $G_1 = \langle (\nabla\rho)^2)\rangle$ and $G_0=\langle |\nabla\rho|\rangle$ with $\langle |\nabla\rho|\rangle = \langle|\nabla\psi|\rangle(\frac{\partial\psi}{\partial\rho})^{-1}$ (where $\langle\cdot\rangle$ denote the flux-surface average).

\subsubsection{Inputs:} The particle fueling rate by pellet injection $\Gamma_{pellet}^i(t)\;\forall i=1,\hdots,n_u$, where $n_u$ is the number of used pellet injectors, comprise the inputs $u(t)\in \mathbb{R}^{n_u}$ to the model\footnote{Note that RAPDENS also contains models for gas fueling and NBI but these are not considered in this work.}. The input vector is defined as
\begin{equation}
u(t) = \begin{bmatrix}\Gamma_{pellet}^1(t), \hdots, \Gamma_{pellet}^{n_{u}}(t)\end{bmatrix}^\top.
\label{eq:RAPDENS input}
\end{equation}

\subsubsection{Nonlinear system of ODEs:} 
The spatial discretization of the density profile (\ref{eq:b-spline discretization}) is inserted in (\ref{appeq:ne PDE})-(\ref{appeq:Nv ODE}) and discretized in time using an equidistant temporal discretization $t_k=t_0+kT_s$ to form the nonlinear set of ODEs
\begin{equation}
x(t_{k+1}) = f(p(t_k),x(t_k),u(t_k)),
\label{eq:nonlinear SS}
\end{equation}
where $f:\mathcal{P}\times\mathbb{R}^{n_x}\times\mathbb{R}^{n_u}\rightarrow\mathbb{R}^{n_x}$ is a nonlinear function\footnote{Note that for the modeling is this paper, the function $f$ is linear with respect to $u$. Hence, (\ref{eq:nonlinear SS}) can be rewritten as $x(t_{k+1}) = f(p(t_k),x(t_k))+B_d(p(t_k))u(t_k)$, where $B_d$ is a parameter varying matrix.} of the state that represents the state evolution due to physical phenomena and the influence of the actuators. 

This formulation is the foundation of the work presented in the coming sections. In sections \ref{sec:DSO} and \ref{sec:EKF results}, it is used to perform state estimation and in sections \ref{sec:MPC method} and \ref{sec:MPC results}, it is used to derive and test a density profile controller.

\section{Dynamic state observer for the density profile}\label{sec:DSO}
A requirement to perform profile control is the estimation of the profile from available measurements. For this purpose, a dynamic state observer (DSO) was proposed in \cite{Blanken2018,Blanken2019}. In this section, we summarize the DSO and we present a forward diagnostic model for TS that will allow us to perform profile estimations using TS measurements. 
\vspace{2mm}
\hrule
\vspace{2mm}
\textbf{Notation 1:} Let a generic variable $e(t)$ be a physical quantity of a real system. The estimate of this quantity made by a DSO at time $t=t_{k_1}$ with diagnostic signals from time $t=t_{k_2}$ is denoted as $\hat{e}_{k_1|k_2}$.
\vspace{2mm}
\hrule
\subsection{Forward Thomson scattering model}\label{ssec:TS model}
The DSO is comprised of a prediction model, a diagnostic model, and an observer gain (see figure \ref{fig:EKF sim setup}). The original observer, developed in \cite{Blanken2018,Blanken2019}, included diagnostic models for interferometry and bremmstrahlung but not for TS. Here, we discuss the inclusion of a TS model that will allow us to perform density estimation with TS measurements.

TS systems are used in tokamaks to measure the radially resolved distributions of electron temperature and density \cite{Gowers2016}. In combination with knowledge of the 2D equilibrium, 
the spatial measurements $n_e(R,Z,t)$ can be transformed into equilibrium measurements of the density $n_e(\rho,t)$ at known equilibrium locations $\bar{\rho}_i$ with $i=1,\hdots,n_{TS}$. $n_{TS}$ denotes the total number of radial measurement locations. The spatial resolution of the ITER core TS system is $67$ mm \cite{Bassan2016}, which, in a typical ITER plasma with a radius of $r_p\approx 2$ m, provides $\approx 29$ measurement locations distributed over the high and low field side. In this work, to be conservative, $n_y=n_{TS}=11$ is chosen.

The TS output vector $y \in \mathbb{R}^{n_{TS}}$ should contain the plasma density evaluated at the corresponding flux labels $\bar{\rho}_i$, i.e.,
\begin{equation} y(t) = \begin{bmatrix} n_e(\rho,t)\Big|_{\bar{\rho}_1}&\cdots& n_e(\rho,t)\Big|_{\bar{\rho}_{n_{TS}}}\end{bmatrix}^\top. \label{eq:physical output equation}\end{equation}
The relation between (\ref{eq:RAPDENS state vector}) and (\ref{eq:physical output equation}) is obtained by evaluating the basis functions used for spatial discretization in (\ref{eq:b-spline discretization}) at $\rho = \bar{\rho}_i\; \forall\; i=1,\hdots,n_{TS}$. Inserting the spatial discretization (\ref{eq:b-spline discretization}) in (\ref{eq:physical output equation}) results in the numerical expression of the output vector as function of the b-spline coefficients:
\begin{equation} y(t) = \begin{bmatrix} \sum_{\alpha=1}^m\Lambda_\alpha(\rho)\Big|_{\bar{\rho}_1} b_\alpha(t) \\ \vdots \\ \sum_{\alpha=1}^m\Lambda_\alpha(\rho)\Big|_{\bar{\rho}_{n_{TS}}} b_\alpha(t) \end{bmatrix}, \label{eq:TS outputs}\end{equation}
where $\Lambda(\rho)|_{\bar{\rho}^i}$ are the basis function evaluated at the corresponding radial locations. By defining $\mathbf{\Lambda_{TS}}\in \mathbb{R}^{n_{TS}\times m}$ the matrix containing the basis function evaluated at all the $n_{TS}$ radial locations, we can formulate an output mapping $C(p(t)) = \begin{bmatrix}\mathbf{\Lambda}_{TS} & 0^{n_{TS}\times 2} \end{bmatrix}$ such that the forward outputs are given by
\begin{equation} y(t) = C\big(p(t)\big)x(t). \label{eq:output mapping}\end{equation}
The output mapping $C$ changes as function of the equilibrium, hence, it depends on the parameter vector $p(t)$ introduced in a previous paragraph. For the simulations presented in this work however, it is assumed for simplicity that the measured locations are constant.

\subsection{Working principles of the observer}\label{ssec:EKF}
A dynamic state observer (DSO) is an algorithm that uses measurements to constrain prediction made by a dynamic model to iteratively estimate the internal state of a system. A brief overview of the working principles of the observer is given here, more details can be found in \ref{app:KF and EKF}.

In the case of the density distribution, we are dealing with a nonlinear system. Hence, we are using the extended Kalman filter (EKF) framework \cite{Anderson1979} (see \ref{appssec:EKF eq}) to estimate the density. The algorithm works as follows. At each time step, the latest measurements $\tilde{y}_k$ are compared with the predicted measurements $\hat{y}_{k|k-1}$ to form the residual $z_k$ (\ref{appeq:KF residual}). The residual is multiplied by the Kalman gain (\ref{appeq:KF gain}) and used to updated the predicted state $\hat{x}_{k|k-1}$ to obtain the state estimate $\hat{x}_{k|k}$. Finally, the predicted state $\hat{x}_{k+1|k}$ and measurement $\hat{y}_{k+1|k}$ at the next time step are computed using RAPDENS (\ref{eq:nonlinear SS}) and the forward diagnostic model (\ref{eq:output mapping}).

The performance, stability, and robustness of the observer depends on the choice of the measurement covariance $R_k$, the process covariance $Q^x_{k}$, and the disturbance covariance $Q^\zeta_{k}$. This is detailed in section \ref{ssec:cov mat}.\\

In this section, we included a forward TS model in a DSO. With this addition, TS measurements can be used to perform estimations of the density profile. In the next section, the performances of the observer are assessed for ITER.


\section{Density profile estimation in ITER baseline simulations using Thomson scattering measurements}\label{sec:EKF results}
In this section, density estimation simulations are discussed where we demonstrate the effectiveness of using TS measurements to update the state predictions in the DSO. 
First, we introduce the simulated plasma scenario and argue the choice of the covariance matrices, i.e. the tuning knobs of an EKF. Subsequently, we discuss the simulation set-up and present the observer's performance in a simulation with plant-observer model mismatch and realistic noise levels.

\subsection{Plasma scenario}\label{ssec:plasma scenario}
The observer is used to estimate the density profile with TS measurements for an ITER H-mode, 15 MA, -5.3 T, baseline discharge. ITER shot 134173 is chosen as reference scenario as it is up to date the most complete prediction of an ITER baseline discharge available. 

\begin{figure}[]
\centering
\includegraphics[scale=1]{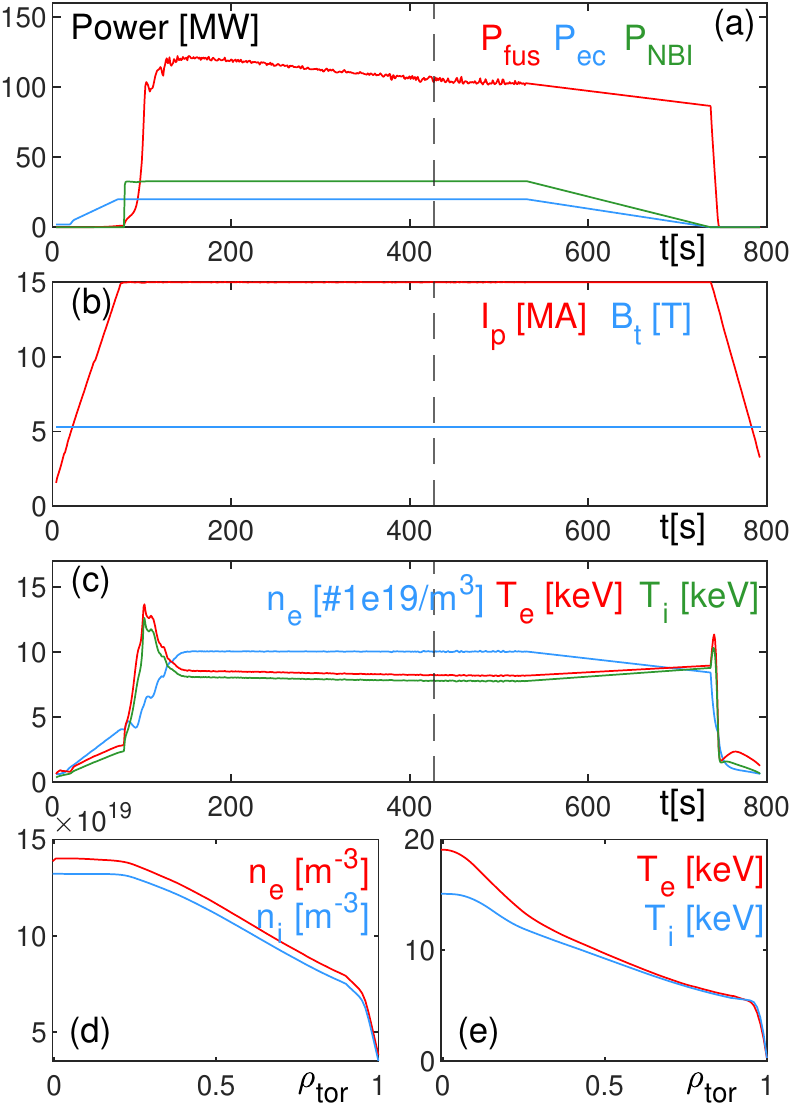}
\caption{Reference scenario ITER \#134173. Evolution of the heating power (a); plasma current and toroidal magnetic field (b); central plasma density and temperature (c); and distributions of density and temperature for $t = 436$ s (d-e). The vertical dashed line indicates the time for which the distributions are shown.}
\label{fig:overview 134173}
\end{figure}

The scenario is obtained with an integrated model (IM) simulation \cite{Koechl2018}. The simulation was performed as an open-loop coupling within IMAS \cite{Imbeaux2015} of the free boundary equilibrium code DINA \cite{Lukash2003,Lukash2005,Lukash2011} with the JINTRAC \cite{Romanelli2014} suite of codes. The baseline scenario was assessed from the formation of the X-point to the transition from X-point to limiter. The time evolution of relevant plasma parameters are given in figure \ref{fig:overview 134173}.(a)-(c). 

In the reference scenario, the ramp-up is modeled to last for 75 s with the L-H transition being triggered by the increase in auxiliary power. The flat-top is modeled to last from $t=75$ s to $t=730$ s. During that time the particle source is pellet injection with a pellet composition of 50\% deuterium and 50\% tritium atoms. It is represented as a continuous source of $2.3\times 10^{22}$ atoms/s deposited at $\rho=0.85$ with Gaussian deposition profile with normalized deposition width of 0.15.

\subsection{Design of covariance matrices}\label{ssec:cov mat}
The accuracy, convergence speed, and robustness of the state estimation depends on the choice of the covariance matrices $R_k$, $Q^x_{k}$, and $Q^\zeta_{k}$ \cite{Anderson1979}. The choice of these matrices is discussed here.

The measurement covariance matrix $R_k$ is chosen as a diagonal matrix with the square root of the modeled noise covariance on the diagonal (the modeled noise is discussed in section \ref{ssec:sim set-up}). Note that this assumes that each TS measurement location is an uncorrelated individual output.

The process covariance matrix $Q_k^{x}$ is constructed as a symmetric Toeplitz matrix\footnote{A \textit{Toeplitz matrix} is a matrix of which the values on the diagonals are constant.} with descending first row. The values on the first row of the Toeplitz matrices determine the spatial correlation between the estimates and is chosen as exponentially decaying. The disturbance covariance matrix $Q_k^\zeta$ is constructed as a diagonal matrix. 

The magnitude of the entries of $Q_k^x$ and $Q_k^\zeta$ reflect the amount of trust in the model predictions and influence the Kalman gain (\ref{appeq:KF gain}) \cite{Anderson1979}. A low observer gain reflects small trust in measurements and high trust in the model predictions. Hence, the EKF will react slowly to the differences between measurements and model predictions. A low gain avoids fitting the noise but relies more heavily on the model predictions, making the reconstruction more sensitive to modeling errors. On the contrary, a high Kalman gain reflects high trust in the measurements and large model uncertainty. The DSO will react fast to discrepancies between model and measurements. The reconstruction will be more robust to modeling error but sensitive to diagnostic noise and errors. 

\begin{figure}[hbtp]
\centering
\includegraphics[scale=.75]{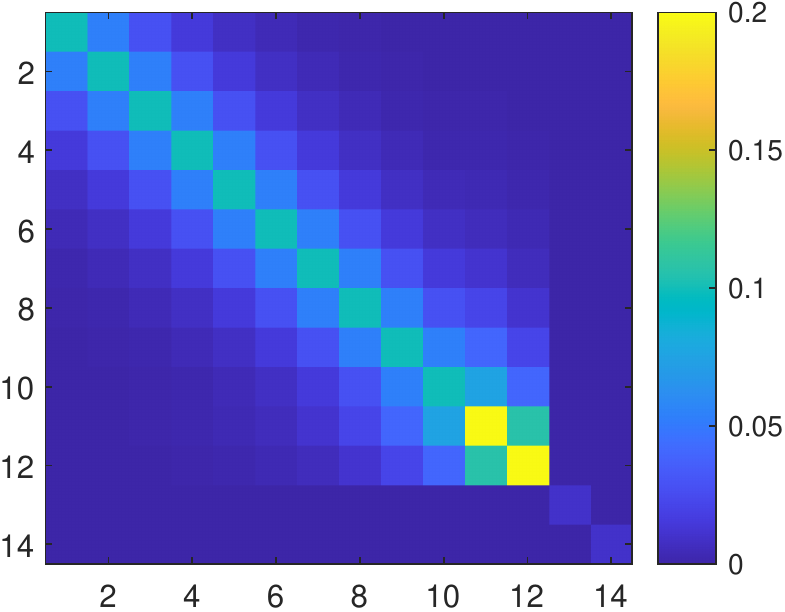}
\caption{Graphical representation of the matrix $Q^x_k$ for the low Kalman gain.}
\label{fig:Covariance}
\end{figure}

The magnitude of the covariance matrices' entries are chosen based on a preliminary analysis of RAPDENS for ITER. The heuristic parameters of RAPDENS are tuned to approximate the flat-top plasma behavior of the reference discharge scenario described in section \ref{ssec:plasma scenario}. From the analysis, it is concluded that the model had good predictive capabilities in flat-top but that modeling of the L-mode and L-H transition is not fully accurate. Hence, a high Kalman gain is chosen in the ramp-up and ramp-down phases and a low Kalman gain is chosen during flat-top. Furthermore, it was observed that the density prediction mismatched at the edge of the plasma. Hence, the diagonal entries of $Q^x_k$ corresponding to the edge of the plasma were increased (see figure \ref{fig:Covariance}).

\subsection{Simulation set-up}\label{ssec:sim set-up}
\begin{figure}[]
\centering
\includegraphics[scale = 1]{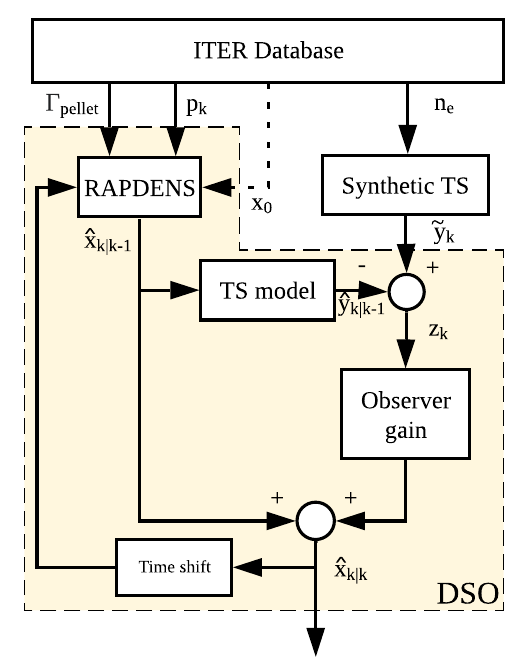}
\caption{Block diagram of the DSO and the set-up used for the density estimation simulations for ITER. The external parameters, the particle fueling rates, and the initial conditions are taken from the ITER database and used to simulate RAPDENS. The 'real' density distribution is used to synthesize synthetic TS measurements. }
\label{fig:EKF sim setup}
\end{figure}

Before proceeding to the estimation results, we briefly describe the set-up of the simulation used to evaluate the observer's performances. A visual depiction of the simulation set-up is given in figure \ref{fig:EKF sim setup}. The EKF is simulated for the entire discharge presented in section \ref{ssec:plasma scenario}. The external parameters required to run RAPDENS (see sec \ref{sec:RAPDENS}), the particle inputs, the initial conditions, and the 'real' (to be estimated) density profile are taken from the database of the IM simulation.

The heuristic parameters of RAPDENS that were tuned to approximate the flat-top plasma behavior in the preliminary analysis are used in the estimation simulation. However, to represent inaccurate knowledge of the transport and investigate the performance of the observer with model uncertainty, systematic plant-observer model mismatch is manually introduced by changing the tuned transport coefficients and the pellet spatial deposition function. 

Furthermore, measurement inaccuracies and noise are included in the simulation with synthetic TS measurements $\tilde{y}_k$. They created by adding artificial noise to the density profile computed by the IM. While the signal-to-noise ratio (SNR) for TS depends on Poisson statistics \cite{Stoneking1997} and thus scales with $1/\sqrt{n_e}$, we choose to model the noise as a Gaussian with expected value $\mathbb{E}[\Delta_{TS}]=0$ and covariance $\sigma_k^2=0.05*max(n_e)$. This way, we approximate the required accuracy of the TS system \cite{Bassan2016} and account for sources of noise such as shot noise, thermal noise in the detection circuit, noise due to plasma light, and noise due to neutron damage on optics installation \cite{Togashi2014,Tanimura1998}. It is important to note that Thomson scattering can also be affected by systematic errors, such as misalignment or due to inaccuracies in the equilibrium reconstruction. These are not accounted for in this work. When using "real" Thomson scattering measurements, detection and correction for these errors can be performed by integrating the Thomson scattering data with other diagnostics in the observer. This is elaborated further in Section \ref{sec:discussion}.

\subsection{Density profile estimation results}\label{ssec:EKF res}
\begin{figure}[]
\centering
\includegraphics[scale=1]{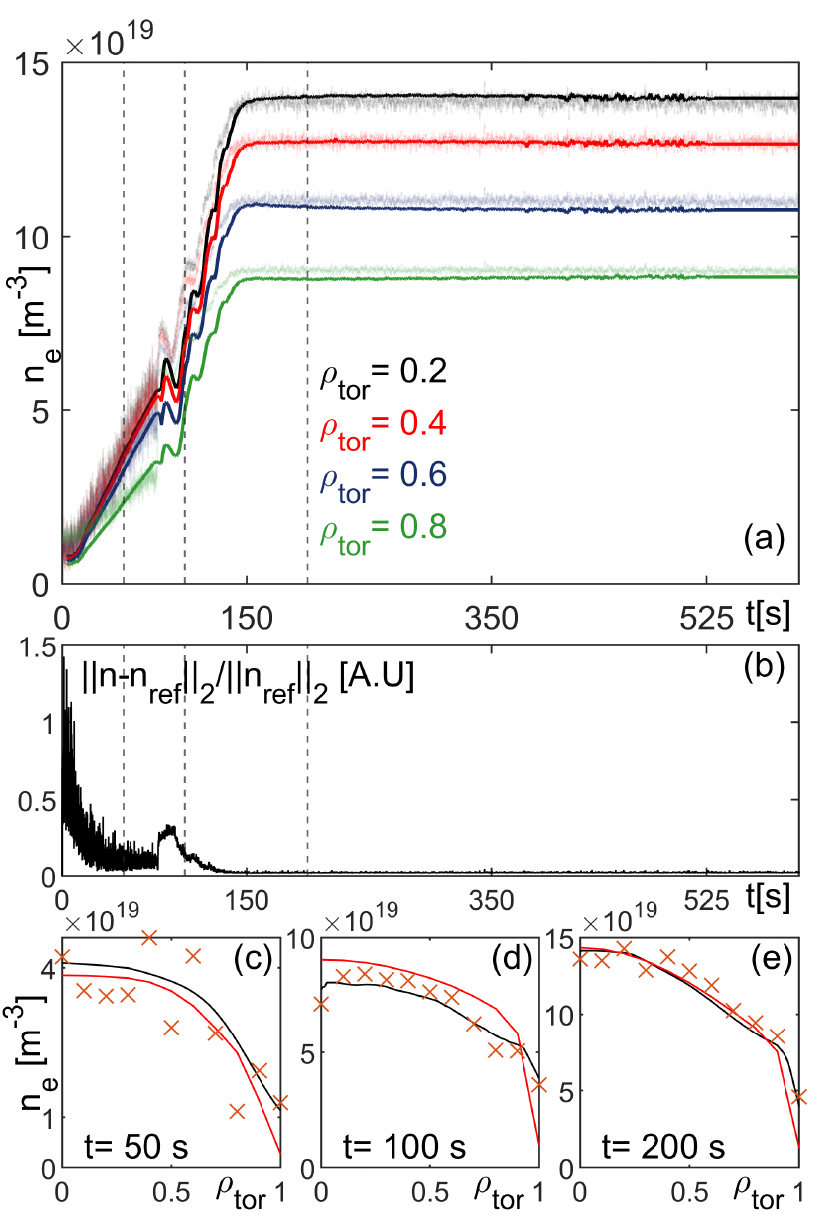}
\caption{Results of density estimation simulations for ITER \#134173 using synthetic TS data to update the state predictions. Plant-observer mismatch is present. Frame a) The estimated density (faded) is compared with the IM density (solid) for four flux labels. Dashed vertical lines show the time at which the profiles are shown. Frame c-e) The estimated density profile (red) is compared with the real density profile (black). The TS measurement points are shown (orange). The observer is capable of accurately estimating the density profile. The estimation is of best quality during the flat-top part of the discharge.}
\label{fig:EKF sim res}
\end{figure}

The simulation results are shown in figure \ref{fig:EKF sim res}. The estimation error, depicted in figure \ref{fig:EKF sim res}.b, is expressed as the normalized 2-norm of the error vector: $||\hat{y}_{k|k}-n_e||_2/||n_e||_2$, where $\hat{y}_{k|k}=C(p_k)\hat{x}_{k|k}$ with $C(p_k)$ defined in (\ref{eq:output mapping}) and $n_e$ the density profile of the IM simulation. Only the part of the profile measured by the core TS system \cite{Bassan2016}, i.e., $0\leq\rho\leq0.9$, are used in the error computation.

Overall, good density reconstruction performances can be seen for the entire profile, i.e., the estimated densities are shown to correctly track the true densities (see figure \ref{fig:EKF sim res}.a). The normalized estimation error averages around  $\approx 10 \%$ during the ramp-up and $\approx2 \%$ for the flat-top (see figure \ref{fig:EKF sim res}.b).

During the ramp-up and the L-H transition, from $t=0$ s to $t=75$ s the effect of the high Kalman gain can clearly be distinguished. During that time, the artificial measurement noise is visible in the estimation, as can be seen in figure \ref{fig:EKF sim res}.b. Decreasing the Kalman gain during this phase would reduce the influence of noise but increase the overall estimation error as the model is not accurate for the early stages of the discharge. 

An increase in the estimation error can be seen between $t=75$ s and $t=90$ s. At $t=75$ s, the plasma enters H-mode. As was discussed in section \ref{ssec:cov mat}, the observer gain is changed from a high to a low Kalman gain at that time. This means that the observer relies more on the model predictions. The modeled density increase in early H-mode is steeper in RAPDENS, hence, the density profile is overestimated (see figure \ref{fig:EKF sim res}.d). This could be solved by slowly transitioning from a high to a low observer gain once the plasma enters H-mode. After $t=90$ s, the estimation error decreases rapidly. 
A good agreement can be seen between the estimated and real density profile for the flat-top (see figure \ref{fig:EKF sim res}.e).

Note that the transport dynamics change after the sudden decrease in the density at $t=634$ s. This change in dynamics is not modeled in RAPDENS and results in an increase of the estimation error to $\approx5\%$.

This section shows that using TS measurements to update the predictions made by the RAPDENS model in an EKF results in accurate density profile estimations. Furthermore, the quality of the estimation is deemed sufficient to be used for profile control, especially during the flat-top.\\


\section{Design of an MPC controller}\label{sec:MPC method}
In this section, the design of an MPC controller for the density distribution is presented. We first introduce the concepts of MPC and motivate the use of this strategy for the control of the density distribution. Subsequently, we discuss in detail each component of the controller.


\subsection{Model predictive control}\label{ssec:MPC}
\begin{figure}[]
 \centering
 \includegraphics[width=\linewidth]{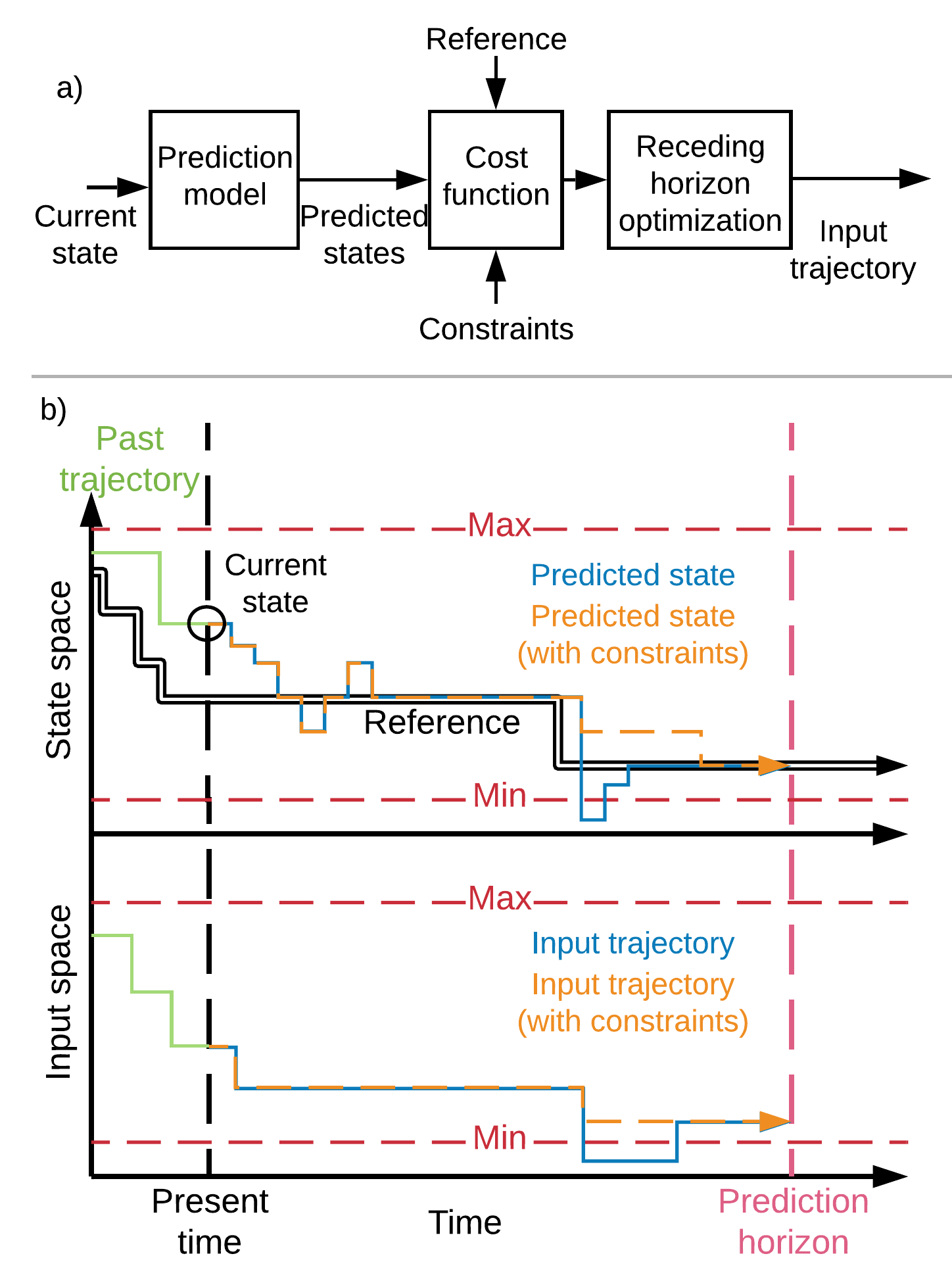}
 \caption{a) Block diagram of the main components of a MPC controller. B) Illustration of MPC for a discrete-time single-input single-output system with constant state and input constraints.}
\label{fig:MPC illustration}
 \end{figure}

Model-predictive control (MPC) is widely adopted as an effective control strategy to deal with multivariate constrained control problems \cite{Qin1997,Lee2011}. A block diagram of an MPC controller is given in figure \ref{fig:MPC illustration}.a.

In such a controller, an explicit model of the to be controlled system is used to predict future states/outputs over a prediction horizon (denoted $N\in\mathbb{N}$). These predictions are used to formulate an optimal control problem where the future tracking error, i.e. the difference between the desired and predicted future outputs is minimized. Limits on the output, states, and inputs can be taken into account as constraints in the optimal control problem. This is illustrated for a discrete-time, single-state single-input system in figure \ref{fig:MPC illustration}.b. 

The design of an MPC controller consists in: choosing a prediction model, formulating a cost function that encompasses the control objectives, formulating constraint functions that translate physical limits of the plasma and actuators as state and input constraints, and implementing and solving the optimal control problem in a numerical optimization solver. These steps are discussed in the coming sub-sections.

\subsection{Control objectives and motivation for MPC}\label{ssec:control objectives}
We aim to synthesize a controller that satisfies the following control objectives:
\begin{enumerate}
\item Track a high-performance reference density profile in a region of interest (ROI).
\item Guaranteeing that the Greenwald density limit is not being violated.
\item Optimizing impurity transport in the ROI.
\end{enumerate}
Note that these objectives are generic, in section \ref{ssec:cont req} the objectives are refined for the specific control simulations.

To control the density profile, the controller will have access to multiple actuators, e.g., multiple pellet injectors \cite{Maruyama2009}, and it will have access to the reconstructed density profile. The control of the density profile is thus a multivariate control problem. Additionally, the Greenwald density and the optimization of impurity transport can be formulated as state and inputs constraints. Hence, the control problem at hand can be classified as multivariate and subject to state and actutator constraints, which is exactly the type of problem for which MPC is suited. 



\subsection{Prediction model}\label{ssec:predmodel}
The first component of a MPC controller is an explicit model of the systems dynamics. In this work, we use a linear disturbance-augmented model derived from RAPDENS for this purpose. 

The prediction model is derived as follows. First, the nonlinear state-space (\ref{eq:nonlinear SS}) is linearized around the desired flat-top conditions $\{x_{ft},p_{ft}\}$ to obtain a linear approximation of the flat-top transport dynamics. Next, as 
the MPC controller runs in discrete-time, the continuous time model is discretized temporally by exact discretization using an equidistant temporal discretization $t_k=t_0+kT_s$, with $T_s = 3$ ms. This results in a discrete-time linear state-space given by
\begin{equation}
\begin{cases} x(t_{k+1})& = Ax(t_k)+ Bu(t_k)\\
y(t_k)& = Cx(t_k).
\end{cases}\label{eq:linear SS disc}
\end{equation}
Next, to incorporate integral action and enable the controller to compensate for steady-state offsets induced by model mismatch and slow changing disturbances, a disturbance model as proposed in \cite{Pannocchia2003a,Maeder2009} is used. To do so, (\ref{eq:linear SS disc}) is augmented to 
\begin{equation}
\begin{cases} x(t_{k+1})& = Ax(t_k)+ Bu(t_k) + B_dd(t_k)\\
d(t_{k+1}) &= d(t_k)\\
y(t_k)&= Cx(t_k)+C_dd(t_k),
\end{cases}\label{eq:LDAM}
\end{equation}
such that it includes the disturbance $d(t)\in \mathbb{R}^{n_d=n_y}$ with disturbance model matrices $B_d$ and $C_d$. These matrices are chosen to guarantee the observability of the augmented system (\ref{eq:LDAM}) \cite{Maeder2009}.

The first condition for the augmented system to be observable is that the non-augmented system is observable. The local linear model (\ref{eq:linear SS disc}) (with state vector (\ref{eq:RAPDENS state vector})) contains two unobservable states. These are the particle inventories of the wall $N_w$ and vacuum $N_v$. Hence, for the controller we use the minimal realization of input-output dynamics. Next, we choose $B_d = I^{n_x,n_d}$ and $C_d =I^{n_d,n_d}$, such that 
\begin{equation} \text{rank}\begin{bmatrix} A-I & B_d\\C & C_d\end{bmatrix}=n_x+n_d. \label{eq:obsv condition}\end{equation}
guaranteeing that the augmented prediction model is observable \cite{Maeder2009}. 

\vspace{2mm}
\hrule
\vspace{2mm}
\textbf{Notation 2:} For a system with state variable $x$ and input variable $u$, we distinguish between the system's state at time $t=t_k$ denoted $x(t_k)$ and the predicted state at time $t=t_k+i$ denoted $x_{i,k}$. Analogously, $u(t_k)$ denotes the input to the system at time $t=t_k$ and $u_{i,k}$ denotes the input that would be applied at time $t=t_k+i$.
\vspace{2mm}
\hrule
\vspace{2mm}

Using notation 2, the discrete-time linear disturbance augmented state-space prediction model is then given by
\begin{equation}
\begin{cases} x_{i+1,k}& = Ax_{i,k}+ Bu_{i,k} + B_dd_{i,k}\\
d_{i+1,k} &= d_{i,k}\\
y{i,k}&= Cx_{i,k}+C_dd_{i,k},
\end{cases}\label{eq:prediction model}
\end{equation}
where $x_{i,k}$, $d_{i,k}$, $y_{i,k}$, and $u_{i,k}$ denoted the predicted state, disturbance, output and control input at $i\in\mathbb{N}$ time steps ahead of the starting time $t=t_k$. 

By introducing, for a given prediction horizon $N$, the stacked notations
\begin{equation}
\underbrace{\bm{X}_k = \begin{bmatrix}
x_{1,k} \\ x_{2,k} \\ \vdots \\ x_{N-1,k}\\x_{N,k}
\end{bmatrix}}_{\text{Future states}}, \; \; \; 
\underbrace{\bm{U}_k = \begin{bmatrix}
u_{0,k} \\ u_{1,k} \\ \vdots \\ u_{N-2,k}\\u_{N-1,k}
\end{bmatrix}}_{\text{Future inputs}},
\label{eq:stacked state and input}
\end{equation}
the compact formulation of the entire prediction sequence is given by
\begin{equation}
\bm{X}_k=\mathbb{A}_xx(t_k)+\mathbb{B}_u\bm{U}_k+\mathbb{B}_dd(t_k),
\label{eq:compact formulation}
\end{equation}
the stacked prediction matrices $\mathbb{A}_x$, $\mathbb{B}_u$, and $\mathbb{B}_d$ are derived in \ref{app:prediction mat}.\\

The prediction model allows us to relate an unknown input sequence $\bm{U}_k$ to the current state $x(t_k)$ and current disturbance $d(t_k)$. In the coming sections, an optimal control problem is formulated using this model that can be used to compute an optimal input sequence. 

\subsection{Disturbance estimation}\label{ssec:dist obsv}
The disturbance $d(t_k)$ introduced in (\ref{eq:LDAM}) is estimated at each time instance using a Kalman filter (KF). The KF equations can be found in \ref{app:KF and EKF}. The performance and stability of the observer (and partially that of the controller) are determined by the choice of the measurements covariance $\mathcal{R}$ and the disturbance covariance $\mathcal{Q}_d$. They have been tuned manually and chosen as diagonal matrices with $\mathcal{R} = 5\times 10^{-3}I_{n_y}$ and $\mathcal{Q}_d= 1\times 10^{-6}I_{n_d}$ 
With this design, the estimate of the disturbances converges in $\approx 2$ s. After convergence, the prediction model is capable of dealing with systematic and slow-varying plant-model mismatches.

\subsection{State and actuator constraints}\label{ssec:constraints}
Next, the construction of the constraints functions is explained. The constraint functions are used to constraint the future states and inputs in the optimization problem. 

Three different type of constraints are imposed on the system in this work: linear input constraints to represent the minimal and maximal fueling rates of the pellet injectors, linear state constraints to represent the Greenwald density limit, and nonlinear state and input constraints to optimize impurity transport.

First, linear constraints are defined to represent actuator limits and plasma density limit. As shown in \ref{app:lin con}, these constraints can be formulated as linear inequality constraints on the input sequence $\bm{U}_k$ as 
\begin{equation} 
A_{\text{ineq}}\bm{U}_k\leq b_{\text{ineq}}.
\label{eq:lin con}
\end{equation}

Second, a nonlinear constraint is imposed to the states and inputs that translate the desire to optimize impurity transport. A condition for outward neo-classical impurity transport can be derived as a favorable ratio between the logarithmic ion temperature gradient $L_{T_i}$ and the logarithmic density gradient $L_{n_e}$, i.e., $L_{T_i}/L_{n_e}\gtrapprox\eta_{ic}\approx 1$  \cite{Angioni2014}. Assuming typical peaked temperature and density profiles in ITER flat-top, the ratio can be used to formulate an inequality constraint on the logarithmic gradient of the density profile:
\begin{align}
 &\frac{|L_{n_e}|}{|L_{T_i}|}\leq 0.95 + \epsilon \nonumber\\
 \equiv & |\frac{\partial n_e(\rho)}{\partial \rho}\frac{1}{n_e(\rho)}|\leq|L_{T_i}|(0.95+\epsilon).
 \label{eq:log const soft}
\end{align}
Note that a soft constraint formulation is used to avoid infeasibility of the optimization problem \cite{Vada1999}. In such a formulation, violation of the constraint is allowed but penalized via the parameter $\epsilon$ and the cost matrix $W_\epsilon$ in the cost function (see (\ref{eq:cost function})). In \ref{app:nonlinear con}, the constraint is formulated as a nonlinear function of the state, control input, disturbance estimate, and soft constraint parameter. The nonlinear constraint function $g_1(x_k,u_k,d_k,\epsilon)$ is defined as (time dependence denoted by subscript $k$ for readability):
\begin{multline}
g_1(x_k,u_k,d_k,\epsilon) \equiv |W_yC'(Ax_k+Bu_k+B_dd_k)|-\\|W_y\Big(C(Ax_k+Bu_k+B_dd_k)+C_dd_k\Big)(0.95W_yL_{T_i}+\epsilon)| \leq0.
\label{eq:nl const}\end{multline}
The linear and nonlinear constraints are used to constrain the optimization problem. With their inclusion the controller accounts for physical limits of the system (actuator and plasma) and the density profile - inward impurity transport interaction. 

\subsection{Cost function}
The objective of the controller is to track the high-performance reference $r_k$ while avoiding aggressive control and violating the constraints. This is expressed in the cost function $J_k$ as:
\begin{equation}
\begin{split}
J_k = \|x_{N,k}-\bar{x}_N\|^2_P+\sum_{j=0}^{N-1}(\|x_{j,k}-\bar{x}_j\|^2_{W_x}+\\\|u_{j,k}-\bar{u}_j\|^2_{W_u})+W_{\epsilon}\epsilon^2,
\end{split}
\label{eq:cost function}
\end{equation}
where for a given matrix $T$ and vector $x$ of appropriate dimensions, we define $\|x\|^2_T\triangleq x^\top Tx$. $\bar{x}_i$ and $\bar{u}_i$ are the desired steady-state states and control inputs, i.e., the states and inputs that are to be reached for the reference to be tracked. They are computed using
\begin{equation} 
\begin{bmatrix}\bar{x}_i \\\bar{u}_i  \end{bmatrix} = \begin{bmatrix} A-I & B \\ HC & 0 \end{bmatrix}^{-1}\begin{bmatrix}
-B_d\hat{d}(t_k) \\ r(t_k)-C_d\hat{d}(t_k)\end{bmatrix}, \label{eq:MPC mathematical problem}
\end{equation}
where $(A,B,B_d,C,C_d)$ are the matrices of (\ref{eq:LDAM}) and $H\in \{0,1\}^{n_z,n_x}$ is the controlled variable matrix. 

In (\ref{eq:cost function}), the \textit{stage state cost} matrix $W_x$ is used to penalize deviations between the desired steady-state state and the future predicted states, up to but excluding the terminal state ($x_{N,k}$). The \textit{stage input cost} matrix $W_u$ is used to weigh the inputs and avoid aggressive control actions. The \textit{terminal state cost} matrix $P$ is used to penalize the deviation of the terminal state. Finally, \textit{soft constraint violation cost} matrix $W_\epsilon$ is used to penalize the violation of the nonlinear state constraint (see section \ref{ssec:constraints}).


An iterative procedure is used to choose the weights. The matrix $W_x$ is chosen as a block-diagonal matrix of two matrices: one being a weighted unity, the other being a zero matrix. The dimensions of these matrices ensure that only the states that parametrize the ROI are penalized. 
The matrix $W_u$ is chosen as a weighted unity matrix, i.e., $W_u=5\cdot10^{-3}I$, and matrix $W_\epsilon$ is chosen as positive definite diagonal matrix with $W_\epsilon\succ Q$. Finally, $P$ is chosen as the solution of the discrete algebraic Riccati equation
\begin{equation} P = A^\top PA-(A^\top PB)(B^\top PB+W_u)^{-1}(B^\top PA)+W_x.  \label{}\end{equation}
By defining these weights, the cost function $J$ is convex and quadratic in $U_k$ \cite{Rawlings2017}.

\subsection{Optimization problem}\label{ssec:opt prb}
The future inputs $\bm{U}_k$ are computed by minimizing the cost function (\ref{eq:cost function}) subject to the constraints (\ref{eq:lin con}) and (\ref{eq:nl const}). Using the prediction model, the nonlinear online optimization is defined as:
\begin{align}
\min_{\bm{U}_k,\epsilon}\;\;\;\; &\bm{U}_k^\top \mathbb{W}_{u,u}\bm{U}_k + \bm{U}^\top_k(\mathbb{W}_{u,x}\hat{x}_{k,k}+\mathbb{W}_{u,d}\hat{d}_{k|k} +\nonumber\\
&\mathbb{W}_{u,r})+W_{\epsilon}\epsilon^2 \nonumber\\
\text{subj.to}\;\;\;\; &A_{\text{ineq}}U_k\leq b_{\text{ineq}},\label{eq:optimal control problem}\\
&g_1(x_k,u_k,d_k,\epsilon)\leq0.\nonumber
\end{align}
The matrices $\mathbb{W}_{u,u}$, $\mathbb{W}_{u,x}$, $\mathbb{W}_{u,d}$, and $\mathbb{W}_{u,r}$ are defined in \ref{app:cost}. The optimization problem is nonlinear due to the presence of the impurity transport constraint (\ref{eq:nl const}). Therefore, sequential quadratic programming \cite{Nocedal2006} is used to solve the problem. Due to the nonlinear constraint, no guarantee on convexity (and global optimality) can be given.


Once an "optimal" solution $\bm{U}_k^*$ of (\ref{eq:optimal control problem}) is obtained, the first $n_u$ entries of the optimal sequence are applied as control inputs to the system. In the next section, the results of control simulations are shown to test the performance of the controller.

\section{Simulation results of closed-loop density distribution control}\label{sec:MPC results}
In this section, we present control simulations that were used to test the MPC controller. Note that it is not the objective of these simulations to provide quantitative estimates of the ITER performance or controllability of a particular scenario, but to illustrate the potential of MPC control for the density profile and help identify challenges that remain to be solved. 

First, we discuss the simulation set-up and refine the control requirements. Next, we demonstrate tracking of a high-performance density profile while minimizing inward impurity transport. This is done in simulations with plant-controller model mismatch and a continuous pellet representation. 

\begin{figure*}[]
\centering
\includegraphics[scale=1]{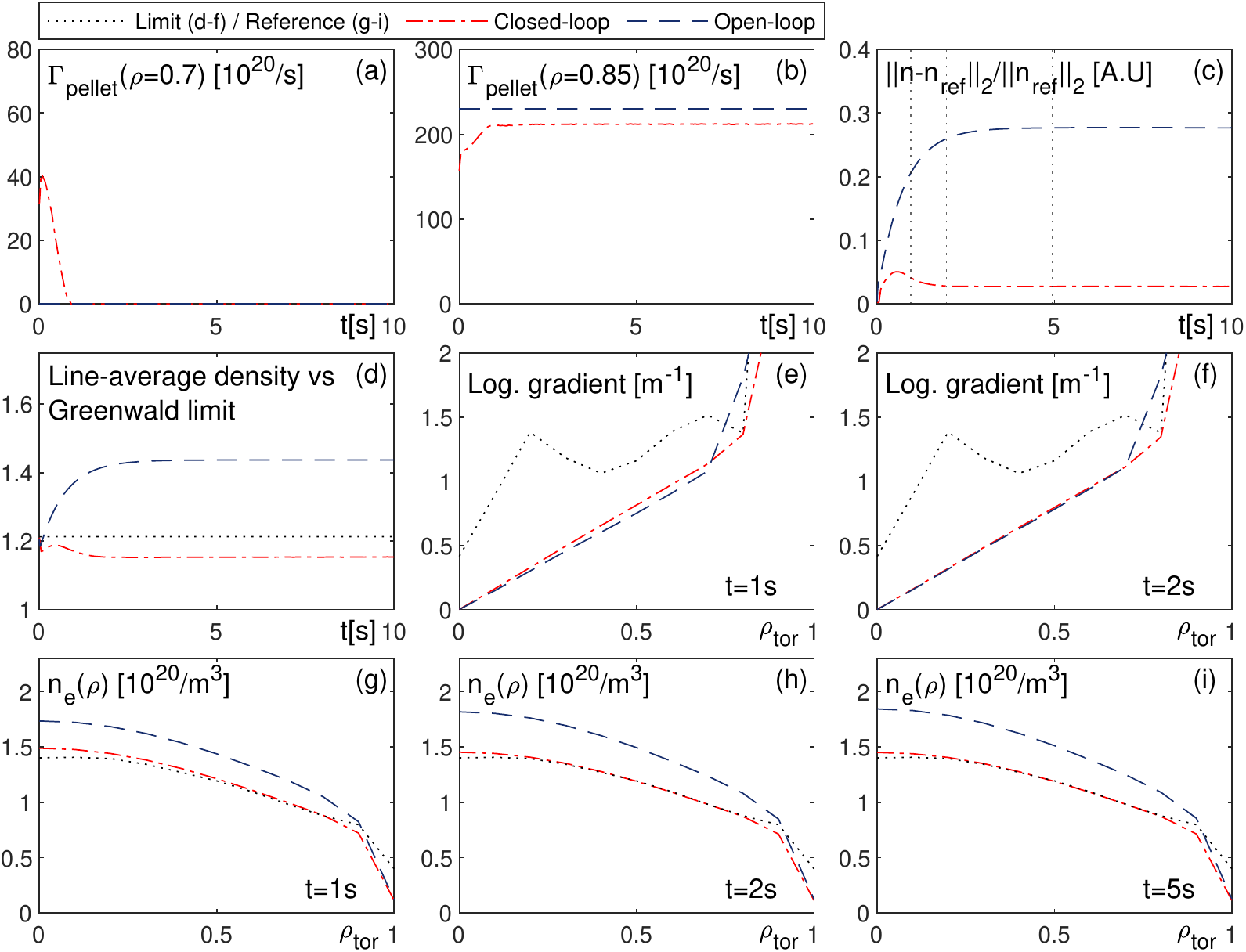}
\caption{Tracking of a high-performance control reference in flat-top phase. The case with feedforward only (blue) is compared with the case with feedback (red). Frame a-b) Required particle flow per pellet injector; Frame c) Relative tracking error for the controlled part of the profile; Frame d) The approximated line average density is compared with the Greenwald density limit (dotted black); Frame e-f) The logarithmic gradient profiles of the ion temperature (dotted black) are shown together with the logarithmic gradients of the density profiles; Frame g-i) Reference density profile (dotted black) is shown together with the controlled profiles for different time instances. The time instances are shown with the dotted vertical lines in c). The MPC controller reduces the tracking error significantly while keeping a favorable ratio between the logarithmic gradient profiles hence minimizing the inward impurity transport.}
\label{fig:CS res1}
\end{figure*}

\begin{figure*}[]
\centering
\includegraphics[scale=1]{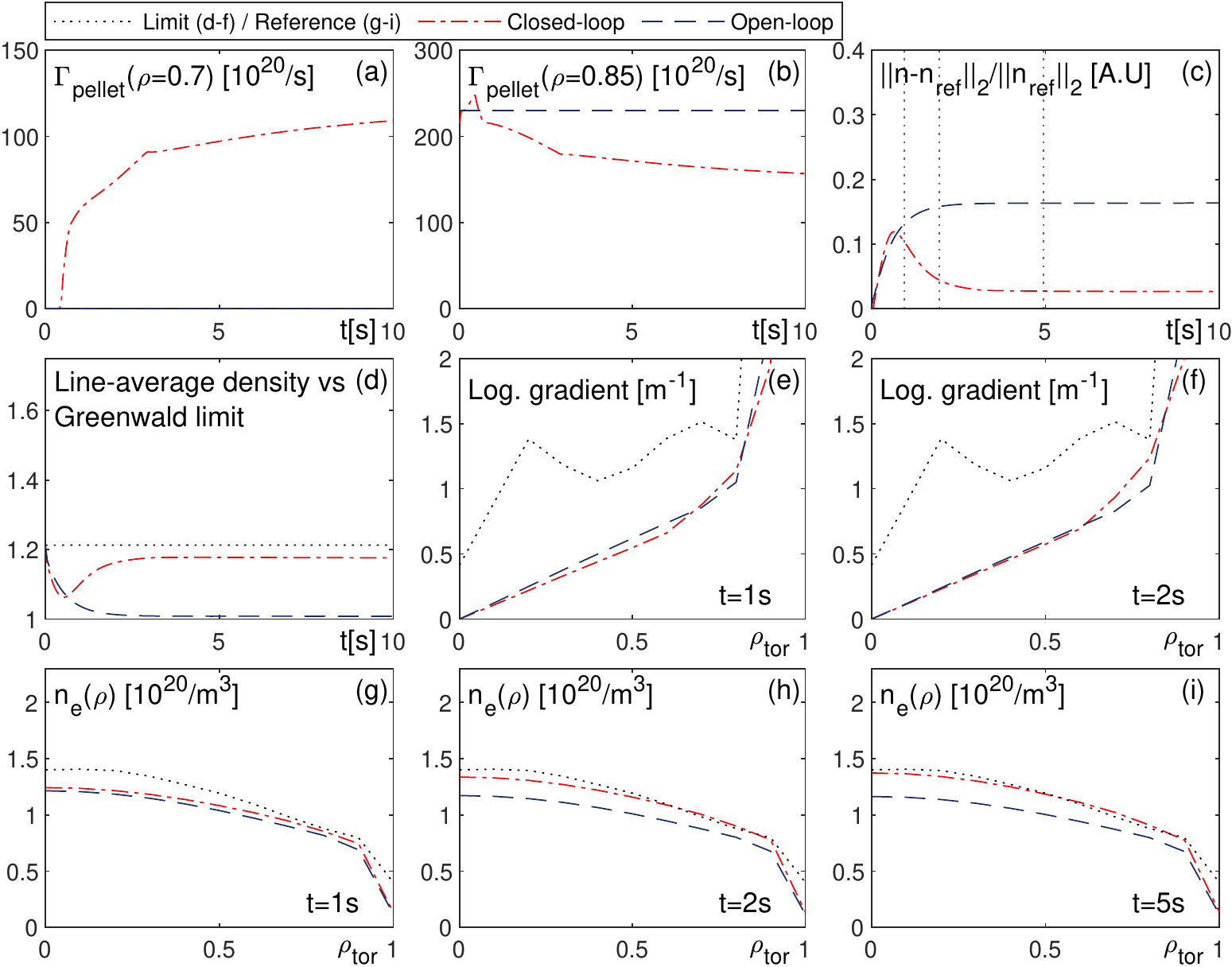}
\caption{Tracking of a high-performance control reference in flat-top phase with uncertain pellet deposition location. The case with feedforward only (blue) is compared with the case with feedback (red). Frame a-b) Required particle flow per pellet injector; Frame c) Relative tracking error for the controlled part of the profile; Frame d) The approximated line average density is compared with the Greenwald density limit (dotted black); Frame e-f) The logarithmic gradient profiles of the ion temperature (dotted black) are shown together with the logarithmic gradients of the density profiles; Frame g-i) Reference density profile (dotted black) is shown together with the controlled profiles for different time instances. The time instances are shown with the dotted vertical lines in c). The MPC controller reduces the tracking error significantly while keeping a favorable ratio between the logarithmic gradient profiles hence minimizing the inward impurity transport.}
\label{fig:CS res2}
\end{figure*}

\subsection{Simulation set-up and control requirements}\label{ssec:cont req}
The considered actuators to perform control are the pellet injection systems that have their injection location situated at the high field side (HFS). During ITER baseline operation, two HFS PIS are planned with the possibility to expand to four \cite{Maruyama2009}. The control inputs (signals changed by the controller) $u(t_k)$ are chosen as the fueling rates of the PIS and defined as 
\begin{equation}
 u(t_k) = \begin{bmatrix}\Gamma_{p,1}(t_k) & \Gamma_{p,2}(t_k)\end{bmatrix}^\top. 
\end{equation}
The deposition function are modeled as parabolic functions with normalized radius $\rho_{p,1}=0.7$ and $\rho_{p,2}=0.85$ and normalized widths of $w_{p,1}=0.3$ and $w_{p,2}=0.15$. These represent obtainable fueling profiles for 3 and 5 mm pellets in ITER \cite{Baylor2007}.

The controlled variables $z(t)$ are the density profile evaluated on an equidistant equilibrium grid for $0\leq\rho\leq0.9$. Similarly, the control reference $r(t)$ is the desired density profile evaluated on the same equilibrium grid. The controller is synthesized using this nominal configuration.

To assess the performance of the controller in simulation, the control objectives discussed in section \ref{ssec:control objectives} are named and specified as follows:
\begin{enumerate}
    \item \textbf{Error (Err) requirement:} The controller can maintain the average tracking error below 5\%. The tracking error is expressed in the normalized 2-norm of the error vector: $||z(t_k)-r(t_k)||_2/||r(t_k)||_2$.
    \item \textbf{Greenwald (Gw) requirement:} The controller can guarantee that the line-average density will not exceed the Greenwald density $n_{GW}$ \cite{Greenwald2002} for any time $t\geq t_0$.
    \item \textbf{Transport (Trsp) requirement:} The controller can optimize neoclassical impurity transport by maintaining a favourable ratio between the logarithmic gradients of the ion temperature and density profiles for the domain $\Omega_c =\{(t,\rho)\in\mathbb{R}\;|\; t\geq t_0,\; 0\leq \rho\leq0.9\}$.
\end{enumerate}

\subsection{Tracking of a reference in flat-top with plant-controller model mismatch}\label{ssec:CS continuous}

The goal of the controller is to compensate for mismatches between the real system  and the model in the controller. Therefore, mismatch is introduced in the plant with respect to the controller model. The results of two simulations are presented here. In the first (Figure \ref{fig:CS res1}), transport mismatch is introduced by decreasing the diffusion coefficient and increasing the drift coefficient. Furthermore, a mismatch is introduced by changing the width of the pellet deposition profile. These changes increase inward transport significantly resulting in higher densities at nominal particle inputs. In the second (Figure \ref{fig:CS res2}), transport mismatch is introduced by increasing the diffusion coefficient and decreasing the drift coefficient in the plant. Furthermore, a mismatch is introduced by changing the deposition radii of the pellet deposition profile, respectively to [0.8,0.9]. These changes decrease inward transport significantly resulting in lower densities at nominal particle inputs.

The open-loop control inputs are computed based on the nominal configuration of the system. For both simulations, the uncertainty in transport results in a large tracking error (figure \ref{fig:CS res1}.c \& \ref{fig:CS res2}.c). In the first simulation, the uncertainty causes a violation of the Greenwald density limit (figure \ref{fig:CS res1}.d) and of the logarithmic gradient constraint (figure \ref{fig:CS res1}.e-f). It is shown in figure \ref{fig:CS res1}.g-i \& \ref{fig:CS res2}.g-i that the open-loop density profiles (blue) do not match the reference profile (black).

The MPC controller is used to control the density profile in closed-loop. The control inputs differ from the open-loop (figure \ref{fig:CS res1}.a-b \& \ref{fig:CS res2}.a-b). It can be seen that in the presented simulations, the controller achieves the control objectives: the tracking error is below $5\%$ (figure \ref{fig:CS res1}.c \& \ref{fig:CS res2}.c); the Greenwald density limit is respected (figure \ref{fig:CS res1}.d); and the favorable ratio between ion-temperature and density logarithmic gradients is maintained (figure \ref{fig:CS res1}.e-f). Furthermore, a good agreement between the controlled (red) and reference (black) profiles can be seen (figure \ref{fig:CS res1}.g-i \& \ref{fig:CS res2}.g-i). 

These results show that with continuous actuators, the MPC controller is capable of tracking a high-performance reference density profile in the presence of plant-controller model mismatch while simultaneously accounting for inward impurity transport.\\ 

Note that the relation between the density and the logarithmic gradient profile depend on a lot of factors, e.g. ratio between transport coefficients, pellet deposition location and deposition profiles, and the fueling rates (control inputs). For the controller to account for inward impurity transport, the control inputs must give enough freedom to find a feasible input sequence that minimizes the control error while respecting the constraint. It is however possible that the control reference and logarithmic gradient constraint become unfeasible (e.g. if the transport coefficients profiles changes drastically), i.e., there does not exist an input sequence for which the logarithmic gradient constraint is respected (this depends greatly on the degrees of freedom of the controller). In this case, the controller cannot optimize the impurity transport. However, it can give a warning to the overarching supervisory controller that the constraint is violated. Subsequently, fast update methods of the transport coefficients \cite{Plassche2020} can be used to update the transport model in the controller and determine a better suited control reference.

\section{Discussion and outlook}\label{sec:discussion}
In this work, we extend a dynamic state observer to estimate the density distribution in ITER using Thomson scattering (TS) measurements. For this, we combine the control-oriented particle transport model RAPDENS with a forward TS model and synthetic TS measurements in an extended Kalman filter (EKF) framework. We have shown that reliable, high-quality density profile estimates can be obtained in simulation with realistic measurement noise levels. However, we use a simple model for the synthetic TS measurements (Gaussian distribution) and thus do not account for systematic errors (e.g., due to equilibrium reconstruction errors or misalignment) or density dependent errors. These error can be present when using real TS measurements. Their presence will reduce the performance of the observer and might require a specific error handling procedure to ensure reliable reconstruction of the density profile. As a next step, we propose to test the extended observer on an experimental reactor with real diagnostic data, preferably AUG or TCV as it was already implemented there \cite{Blanken2018}. On these reactors, the TS measurements will be used together with other diagnostics (interferometers on TCV; interferometers and radiation measurements on AUG) to reconstruct the density profile. It can then be investigated if the expected density estimation performances are achieved and the increase in reconstruction performance by including TS can be quantified. Testing the observer with experimental data will help identify if additional handling is required to deal with systematic errors and design it if need be.

Furthermore in this work, we have synthesized a model-predictive controller for the density distribution in a tokamak. The controller uses the pellet fueling rate as control input to track a high-performance reference profile, avoid density limits, and optimize for impurity transport. We show that for continuous actuators, the controller is capable 
of achieving these control requirements in simulations with plant-controller model mismatch.

Even so, we use the heuristic control-oriented model RAPDENS to synthesize and validate the controller. For further validation, it will be relevant as a next step to couple the controller to more complex first-principles transport codes, e.g., to JETTO as part of JINTRAC \cite{Romanelli2014}  to perform closed-loop control simulations, and investigate the achieved control performances. 
Additionally, pellets are inherently discrete events. However since this works presents a for the first time a controller for the density profile, this discrete nature is not included. Knowing that in a real reactor, the pellet size cannot be changed for each control action and is subject to constraints linked to for example plasma penetration, it will be relevant to investigate the performance of the controller when the discrete pellet nature is introduced in the control loop and when only pellets of a fixed size can be used.

Furthermore, we would like to address the possible extensions of the choice of control inputs. In this work and in literature, e.g. \cite{Ravensbergen2018,Lang2018}, the pellet fueling rate is considered as control input. This is a straightforward choice as it enables the use of linear controllers. In practice, however, the actuation with pellet injection is a complex function of pellet size, pellet velocity, injection frequency, and plasma temperature. 
We think it is relevant to study the possibility of using a more complete input space, e.g., by also taking into account pellet velocity, and analyze the effect on controllability and control performances.

\vspace{2mm}\textbf{Acknowledgments}
This work has been carried out within the framework of the EUROfusion Consortium and has received funding from the Euratom research and training programme 2014-2018 and 2019-2020 under grant agreement No 633053.
The views and opinions expressed herein do not necessarily reflect those of the European Commission.
The authors would like to thank T.Ravensbergen for his help with Simulink and the many great discussion that helped the work forward.
\section*{References}
\bibliographystyle{iopart-num}
\bibliography{references}

\begin{appendix}
\section{Summary of RAPDENS equations}\label{app:RAPDENS}
In this appendix, a summary of the RAPDENS equations is given. 
Mass conservation and radial transport \cite{Hinton1976} are used to model the evolution of the radial plasma density $n_e(\rho,t)$ as
\begin{equation} \left(\frac{\partial V}{\partial \rho}\right)^{-1} \frac{\partial}{\partial t}\left(n_{\mathrm{e}} \frac{\partial V}{\partial \rho}\right)+\left(\frac{\partial V}{\partial \rho}\right)^{-1} \frac{\partial \Gamma_{\mathrm{e}}}{\partial \rho}=S_{\mathrm{e}} \label{appeq:ne PDE}\end{equation}
where $\rho$ is the normalized toroidal magnetic flux, $V$ is the plasma volume, $\Gamma_e(\rho,t)$ is the radial transport flux modeled as
\begin{equation} \Gamma_{\mathrm{e}}=-\frac{\partial V}{\partial \rho}\left(\left\langle(\nabla \rho)^{2}\right\rangle \chi \frac{\partial n_{e}}{\partial \rho}+\langle|\nabla \rho|\rangle \nu n_{e}\right) \label{appeq:Gamma transp}\end{equation}
with $\chi(\rho,c_{HL})$ the diffusivity and $\nu(\rho,c_{LH},I_p)$ the pinch (or drift) velocity where $c_{LH}$ indicates the regime of the plasma (L or H-mode). In \ref{appeq:Gamma transp}, $S_e(\rho,t)$ is the net electron source modeled as
\begin{equation} S_{\mathrm{e}}=\langle\sigma v\rangle_{\mathrm{iz}} n_{n} n_{e}-\langle\sigma v\rangle_{\mathrm{rec}} n_{e} n_{i}-\frac{\left.n_{e}\right|_{\rho}>1}{\tau_{\mathrm{SOL}}}+S_{\mathrm{pellet}}+S_{\mathrm{NBI}} \label{appeq:Particle source}\end{equation}
with $\langle\sigma v\rangle_{\mathrm{iz}}(T_e)$ and $\langle\sigma v\rangle_{\mathrm{rec}}$ the temperature dependent cross-sections for ionization and recombination \cite{Wesson2011}, $\tau_{SOL}$ the time constant for particle loss in the scrape-off layer, $S_{pellet}$ the net particle source due to pellet injection, and $S_{NBI}$ the net particle source due to neutral beam injection. 

The evolution of the particle inventories $N_w(t)$ in the wall and neutral vacuum density $n_n(t)$ are modeled by the 0D ODE's
\begin{equation} \frac{d N_{w}}{d t}=-\frac{N_{w}-c_{w} V_{v, 0} n_{\mathrm{n}}}{\tau_{\text {release }}}+\left(1-\frac{N_{\mathrm{w}}}{N_{\text {sat }}}\right) \int_{V_{\text {SOL}}} \frac{n_{e}}{\tau_{\text {SOL }}} \mathrm{dV}, \label{appeq:Nw ODE}\end{equation}
\begin{equation} \begin{aligned}
V_{\mathrm{v}} \frac{d n_{\mathrm{n}}}{\mathrm{dt}}=& \frac{N_{\mathrm{w}}-c_{w} V_{\mathrm{v}, 0} n_{\mathrm{n}}}{\tau_{\text {release }}}+\frac{N_{\mathrm{w}}}{N_{\text {sat }}} \int_{V_{\mathrm{SOL}}} \frac{n_{e}}{\tau_{\mathrm{SOL}}} \mathrm{dV} \\
&+\int_{V_{\mathrm{p}}}\left(\langle\sigma v\rangle_{\mathrm{rec}} n_{e} n_{i}-\langle\sigma v\rangle_{\mathrm{iz}} n_{n} n_{e}\right) \mathrm{dV} \\
&-\frac{n_{\mathrm{n}} V_{\mathrm{v}, 0}}{\tau_{\mathrm{pump}}}+\left.\Gamma\right|_{\mathrm{\rho}_{\mathrm{e}}}+\Gamma_{\mathrm{valve}},
\end{aligned} \label{appeq:Nv ODE}\end{equation}
with $\tau_{release}$ the particle release rate, $N_{sat}$ the wall saturation inventory, $\tau_{pump}$ the pumping time scale, $c_w$ a dimensionless constant that determines the steady-state balance between the wall and vacuum inventory, and $V_{v,0}$ the nominal vacuum volume.

\section{Dynamic state observer: theory and application to density profile estimation}\label{app:KF and EKF}
\subsection{Kalman filter equations}\label{appssec:KF eq}
Given a system $G$ with state vector $x$, output vector $y$, and input vector $u$, the Kalman filter (KF) assumes that the dynamics of $G$ evolve in discrete-time following a linear state-space as
\begin{equation}
\begin{cases} x(t_{k+1})& = \textbf{F}(t_k)x(t_k)+ \textbf{B}(t_k)u(t_k)+w(t_k)\\
y(t_k)& = \textbf{H}(t_k)x(t_k)+\textbf{D}(t_k)u(t_k)+v(t_k),
\end{cases}\label{eq:KF linear SS}
\end{equation}
where the state transition dynamics are captured in matrix $\textbf{F}$ and the input dynamics in $\textbf{B}$. The output model and influence of the inputs on the outputs are captured in matrices $\textbf{H}$ and $\textbf{D}$. The stochastic behavior of measurement and process noises are modeled in $v$ and $w$ as zero-mean Gaussian white noises with respective covariance matrices $\textbf{R}$ and $\textbf{Q}$ such that $v(t_k)\sim \mathcal{N}(0,\textbf{R}(t_k))$ and $w(t_k)\sim \mathcal{N}(0,\textbf{Q}(t_k))$. The tuning knobs of the KF are the covariance matrices $\mathbf{R}$ and $\mathbf{Q}$.

The algorithm is initialized with an initial state estimate $\hat{x}_{1|0}$ (with corresponding output estimate $\hat{y}_{1|0}$) and an initial \textit{a priori} estimate covariance $\Sigma_{1|0}$. For each discrete time instance $t=t_k$, the estimate of the states are obtained as follows (for readability the time dependence is denoted by the subscript $k$). \\

\noindent\textbf{Update step:}\\
Using the latest available measurements $\tilde{y}_k$, the residual $z_k$ and innovation covariance $\Omega_k$ are computed by
\begin{align}
z_k &= \hat{y}_{k|k-1}-\tilde{y}_k\label{appeq:KF residual}\\
\Omega_k &= \textbf{H}_k\Sigma_{k|k-1}\textbf{H}_k^\top+\textbf{R}_k\label{eq:meas cov},
\end{align}
where $\textbf{H}(t_k)$ is the output matrix and $\Sigma_{k|k-1}$ the \textit{a priori} process covariance. Next, the \textit{optimal} Kalman gain $L_k$ is computed with
\begin{equation}
L_k = \Sigma_{k|k-1}\textbf{H}_k^\top\Omega_k^{-1}, \label{appeq:KF gain}
\end{equation}
and is used to obtain the updated state estimate $\hat{x}_{k|k}$ and the \textit{a posteriori} state estimate covariance $\Sigma_{k|k}$ as
\begin{align}
\hat{x}_{k|k}&=\hat{x}_{k|k-1}+L_kz_k,\label{eq:KF update}\\
\Sigma_{k|k} &= (I-\Sigma_{k|k-1}\textbf{H}_k^\top\Omega_k^{-1}\textbf{H}_k)\Sigma_{k|k-1}.\label{eq:KF aposteriori cov}
\end{align}

\noindent\textbf{Prediction step:}\\
Based on the most likely state $\hat{\textbf{x}}_{k|k}$, the actuator inputs at the present time $\textbf{u}_k$, and the process covariance $\textbf{Q}$, the one step ahead predicted state $\hat{\textbf{x}}_{k+1|k}$ and outputs $\hat{\textbf{y}}_{k+1|k}$ and the \textit{a priori} covariance at time step $t=t_{k+1}$ are given by
    \begin{align}
        \hat{x}_{k+1|k} &= \textbf{F}_k\hat{x}_{k|k}+\textbf{B}_k\textbf{u}_{k}\label{eq:KF pred state}\\
        \hat{y}_{k+1|k} &= \textbf{H}_k\hat{x}_{k|k}+\textbf{D}_k\textbf{u}_{k}
        \label{eq:KF pred output}\\
        \Sigma_{k+1|k} &= \textbf{F}_k\Sigma_{k|k}\textbf{F}_k^\top+\textbf{Q}_k\label{eq:KF apriori cov}.
    \end{align} 

\subsection{Extended Kalman filter equations}\label{appssec:EKF eq}
In case a system cannot be represented accurately by a linear model, the EKF equations are used. The observer assumes that the system's dynamics evolve following
\begin{equation}
\begin{cases} x(t_{k+1})& = \textbf{f}(x(t_k),u(t_k))+w(t_k)\\
y(t_k)& = \textbf{h}(x(t_k),u(t_k))+v(t_k),
\end{cases}\label{eq:EKF nonlinear SS}
\end{equation}
where $\textbf{f}:\mathbb{R}^{n_x}\times\mathbb{R}^{n_u}\rightarrow\mathbb{R}^{n_x}$ and $\textbf{h}:\mathbb{R}^{n_x}\times\mathbb{R}^{n_u}\rightarrow\mathbb{R}^{n_y}$ are non-linear functions of the state and input. The stochastic variables $v$ and $w$ are similar as defined in \ref{appssec:KF eq}.

There are two main differences between the KF and EKF equations:
\begin{enumerate}
\item The matrices $\textbf{F}$ and $\textbf{H}$ (used in the computation the measurement covariance (\ref{eq:meas cov}), the Kalman gain (\ref{appeq:KF gain}), and the state covariances (\ref{eq:KF apriori cov}) and (\ref{eq:KF aposteriori cov})) are the jacobians of $\textbf{f}$ and $\textbf{h}$
, i.e.,
\begin{align}
    \textbf{F}(t_k)& = \frac{\partial\textbf{f}}{\partial x}\bigg\rvert_{x_{k-1|k-1}}& \textbf{H}(t_k)& = \frac{\partial\textbf{h}}{\partial x}\bigg\rvert_{x_{k-1|k-1}}.
\end{align}  
\item The prediction step is performed using the nonlinear model (\ref{eq:EKF nonlinear SS}), i.e., (\ref{eq:KF pred state}) and (\ref{eq:KF pred output}) become
\begin{align}
 \hat{x}_{k+1|k} = \textbf{f}(\hat{x}_{k|k},u_k)\label{eq:EKF pred state}\\
 \hat{y}_{k+1|k} = \textbf{h}(\hat{x}_{k|k},u_k)\label{eq:EKF pred output}
\end{align}
\end{enumerate} 
  
It is important to note that the EKF is a linearized version of the Kalman filter for nonlinear dynamical systems, hence, no guarantee about stability or estimation accuracy can be given \cite{Anderson1979}. These properties are to be checked experimentally for the specific implementations.

\subsection{Application of the in the DSO for the density profile estimation}\label{appssec:DSO ne profile}
For the density reconstruction, the RAPDENS plasma simulator is used as the one step ahead prediction model. The physical state vector $x(t_k)$ (\ref{eq:RAPDENS state vector}) is augmented with additive unknown disturbances $\zeta(t_k) \in \mathbb{R}^{n_x}$. These disturbances are co-estimated by the observer and give a measure of the modeling errors, unmodeled processes, unaccounted particles sources, and errors in the diagnostics models. The 
stochastic behavior of the state and disturbances is modeled by additive zero-mean white noises $w^x_{k}$ and $w^\zeta_{k}$ with respective covariance matrices $Q_k^x$ and $Q_k^\zeta$. The augmented nonlinear state-space model is given by
\begin{equation}
\begin{bmatrix}x_{k+1}\\\zeta_{k+1}\end{bmatrix} = \begin{bmatrix} f_d(x_k,p_k) & B_{\zeta}\zeta_k\\0 & \zeta_k \end{bmatrix}+\begin{bmatrix}B_d(p_k)\\0\end{bmatrix}u_k+\begin{bmatrix}w^x_k\\w^\zeta_k\end{bmatrix}
\label{eq:augm EKF system}
\end{equation}
where $B_\zeta=\begin{bmatrix}I^{m\times m} & 0^{m\times 2}\end{bmatrix}$.\\ An augmented state is defined as
\begin{equation}
\xi_k=\begin{bmatrix}x_k & \zeta_k\end{bmatrix}^\top.
\end{equation}
The stochastic measurement noise is represented by an additive zero-mean white noise $v_k$ with associated covariance matrix $R_k$. The outputs are assumed to evolve as:
\begin{equation}
y_{k}=C(p_k)x_k+ v_k.
\end{equation}
The final step consists in defining the matrices
\begin{align}
    F_k &= \begin{bmatrix} \frac{\partial f}{\partial x_k}\big\rvert_{\hat{\textbf{x}}_{k|k},p_k} & B_\zeta\\0 & I^{m\times m} \end{bmatrix} & B_k &= \begin{bmatrix} B_d \\ 0 \end{bmatrix} \nonumber\\
    H_k&=\begin{bmatrix} C(p_k) & 0 \end{bmatrix}.\label{appeq:EKF matrices RAPDENS}
\end{align}
The augmented state $\xi(t_k)$ is then estimated by using (\ref{eq:augm EKF system}) and (\ref{appeq:EKF matrices RAPDENS}) in (\ref{appeq:KF residual})-(\ref{eq:KF apriori cov}).

\section{Controller: definitions and derivations}
In this appendix we provide the definitions and derivations of the controller matrices described in section \ref{sec:MPC method}.

\subsection{Prediciton model matrices}\label{app:prediction mat}
For a system described by a linear time-invariant disturbance augmented model such as (\ref{eq:LDAM}), the predicted future states $x_{i,k}\;\text{for } i=0:N$ can be related to the current state $x_k$, current disturbance estimate $\hat{d}_k$, and input sequence $u_{k},\hdots,u_{k+N-1}$ by:
\begin{equation}
x_{i,k} = A^ix_{0,k}+\sum_{j=0}^{i-1}A^{j}Bu_{i-1-j,k}+\sum_{j=0}^{i-1}A^{j}B_d\hat{d}_k.
\label{eq:predicted state rewritten}
\end{equation}
We define matrices $\mathbb{A}_x$, $\mathbb{B}_u$, and $\mathbb{B}_d$ as
\begin{equation}
\mathbb{A}_x = \begin{bmatrix}
A \\A^2 \\ \vdots \\ A^{N-1} \\ A^{N}
\end{bmatrix},
\label{eq:PHI}
\end{equation}
\begin{equation}
\mathbb{B}_u = \begin{bmatrix}
B & 0 & \cdots & 0 \\
AB & B & \cdots& 0 \\
\vdots & \vdots & \ddots & \vdots \\
A^{N-1}B & A^{N-2}B & \cdots& B
\end{bmatrix},
\label{eq:GAMMA}
\end{equation}
\begin{equation} \mathbb{B}_d = \begin{bmatrix}
B_d & 0 & \cdots & 0 \\
AB_d & B_d & \cdots& 0 \\
\vdots & \vdots & \ddots & \vdots \\
A^{N-1}B_d & A^{N-2}B_d & \cdots& B_d
\end{bmatrix}\begin{bmatrix}
I \\ \vdots \\ \vdots \\ I
\end{bmatrix}.
\label{eq:GAMMA dist}
\end{equation}
Using (\ref{eq:PHI}), (\ref{eq:GAMMA}), and (\ref{eq:GAMMA dist}), we rewrite (\ref{eq:predicted state rewritten}) for the stacked predicted states and input sequence introduced in notation 2:
\begin{equation}
\bm{X}_k=\mathbb{A}_x\hat{x}_k+\mathbb{B}_u\bm{U}_k+\mathbb{B}_d\hat{d}_k.
\label{eq:Xk}
\end{equation}

\subsection{Linear state and actuator constraints}\label{app:lin con}
Here the linear and nonlinear constraints are derived. The pellet particle input is constrained as $0\leq\Gamma_{pellet}\leq\Gamma_{pellet}^{max}$. The pellet injectors are designed to provide a mass flow rate $\dot{m}=0.23+30\%=0.33\text{ gs}^{-1}$ of solid DT to the plasma \cite{Baylor2007}. The maximum electron fueling rate is than given by
\begin{equation} \dot{\Gamma}_{pellet}^{max}=\dot{m}*(f_DM_D+f_TM_T)*N_{avg}, \label{}\end{equation}
where $f_D$ and $f_T$ are the fractions of deuterium and tritium atoms in the pellet, $M_D$ and $M_T$ the molecular masses of the two isotopes and $N_{avg}$ representing the Avogadro constant. Assuming a pellet composed at 50\% of deuterium and 50\% of tritium, the maximal fueling rate is equal to $\dot{\Gamma}_{pellet}^{max}=4.115e22\;[\#/s]$. Hence, the number of injected particles is constraint by $\Gamma_{pellet}^{max} = 2.57e21\;[\#]$.

No lower constraint exist for the states hence $x_{min}=-\infty$. The Greenwald density limit \cite{Greenwald2002} sets an upper bound on the line-average electron density in a specific device configuration. This upper limit is given by 
\begin{equation}
    n_{gw} = \frac{I_p}{\pi a^2}
    \label{eq:ch2.ngreenwald}
\end{equation}
where $I_p$ is the plasma current and $a$ the minor radius of the tokamak. By constructing a matrix $Z\in\mathbb{R}^{1\times n_x}$ such that $Zx\approx \bar{n}$ where $\bar{n}$ is the line-average density, this limit can be formulated as a linear state constraint.

The linear constraints are given by:
\begin{align} 
u_{min} \leq u_{i,k} \leq u_{max}, &\; \forall i=0,1,\cdots,N-1,\nonumber\\
x_{min} \leq x_{i,k} &\; \forall i=0,1,\cdots,N,\nonumber\\ 
Zx_{i,k} \leq n_{gw} &\; \forall i=1,2,\cdots,N,
\label{eq:linear constraints form 1}
\end{align}
By defining
\begin{align}
&M_i=\begin{pmatrix} 0_{n_u\times n_x}\\0_{n_u\times n_x}\\-I_{n_x}\\Z\end{pmatrix},\;\;E_i=\begin{pmatrix} -I_{n_u}\\I_{n_u}\\0_{n_x\times n_u}\\0_{1\times n_u}\end{pmatrix},\;\;b_i=\begin{pmatrix}-u_{min}\\u_{max}\\-x_{min}\\n_{gw}\end{pmatrix}\nonumber\\
&M_N=\begin{pmatrix}-I_{n_x}\\Z\end{pmatrix},\;\;\text{and } b_N=\begin{pmatrix}-x_{min}\\n_{gw}\end{pmatrix},
\end{align}
we rewrite (\ref{eq:linear constraints form 1}) as
\begin{align}
&M_ix_{i,k}+E_iu_{i,k}\leq b_i\;\;\forall i=0,1,\hdots,N-1\nonumber\\
&M_Nx_{N,k}\leq b_N.
\label{eq:linear constraints form 2}
\end{align}
Using the stacked notation (\ref{eq:stacked state and input}) and defining $\mathcal{M}_0$, $\mathcal{M}_i$, $\mathcal{E}_i$ and $\mathbf{b}$ as
\begin{equation}
\mathcal{M}_0=\begin{pmatrix}
M_0\\ 0 \\\vdots \\ 0
\end{pmatrix},
\label{eq:mathcal M0}
\end{equation}
\begin{equation}
\mathcal{M}_i=\begin{pmatrix}
0 & \cdots & 0 \\
M_1 & \cdots & 0\\
\vdots & \ddots & \vdots\\
0 & \cdots & M_N
\end{pmatrix},
\label{eq:mathcal Mi}
\end{equation}
\begin{equation}
\mathcal{E}_i=\begin{pmatrix}
E_0 & \cdots & 0 \\
\vdots & \ddots & \vdots\\
0 & \cdots & E_{N-1}\\
0 & \cdots & 0
\end{pmatrix},
\label{eq:mathcal Ei}
\end{equation}
\begin{equation}
\mathbf{b}=\begin{pmatrix}
b_0 \\ b_1\\ \vdots\\b_n
\end{pmatrix},
\label{eq:mathcal b}
\end{equation}
the constraints in (\ref{eq:linear constraints form 2}) are grouped as
\begin{equation}
\mathcal{M}_0x(k)+\mathcal{M}_i\bm{X}_k+\mathcal{E}_i\bm{U}_k\leq\mathbf{b}.
\label{eq:lin con form 3}
\end{equation}
Finally, the matrices of (\ref{eq:lin con}) are obtained by inserting (\ref{eq:Xk}) in (\ref{eq:lin con form 3}) and are defined as:
\begin{align}
A_{ineq}&=\mathcal{M}_i\Gamma_u+\mathcal{E}_i,\\
b_{ineq}&=\mathbf{b}-(\mathcal{M}_0+\mathcal{M}_i\Gamma_x)x(k)-\mathcal{M}_i\Gamma_dd(k).
\end{align}

\subsection{Nonlinear state constraints}\label{app:nonlinear con}

Here we formulate the nonlinear constraint function $g_1$ introduced in section \ref{ssec:constraints}.
Recalling that $y(t_{k+1})=n_e(\bar{\rho},t_{k+1})$, with $\bar{\rho}$ defined in section \ref{ssec:TS model}, the augmented model (\ref{eq:LDAM}) can be used to relate the density profile at time $t=t_{k+1}$ can be related to the state and inputs at time $t=t_k$. The relation is given by:
\begin{align}
n_e(\bar{\rho},t_{k+1})&= y(t_{k+1})\nonumber\\
 &= Cx(t_{k+1})+C_dd(t_{k+1})\nonumber\\
  &= C(Ax(t_k)+Bu(t_k)+B_dd(t_k))+\nonumber\\
  & \;\;\;\;\;C_dd(t_{k+1}). \label{eq:ne to x and u 1}
\end{align}
Under the assumption of slow varying disturbances, such that $d(t_{k+1})\approx d(t_k)$, we can write
\begin{equation} n_e(\bar{\rho},t_{k+1})= C(Ax(t_k)+Bu(t_k)+B_dd(t_k))+C_dd(t_{k}).\label{eq:ne to x and u 2}\end{equation}
Finally, we define $C^\prime$ as
\begin{equation} C^\prime = \frac{\partial\Big(\sum_{\alpha=1}^m\Lambda_\alpha(\bar{\rho})\Big)}{\partial\rho}, \label{eq:C prime}\end{equation}
such that $C^\prime x(t_k)=\frac{\partial n_e(\rho,t_k)}{\partial \rho}$ and we construct an output weight matrix $W_y$ to weight the importance of the outputs in the nonlinear constraint. In this work we chose this matrix such that the weight of the outputs for which $\rho<0.9$ are unity and the outputs for which $\rho>0.9$ are weighted 0. 

Inserting (\ref{eq:ne to x and u 2}), (\ref{eq:C prime}), and $W_y$ in (\ref{eq:log const soft}) and rewriting the equation in negative null form, we can formulate the nonlinear constraint function $g_1$ as (time dependence denoted by subscript $k$ for readability):
\begin{multline}
g_1(x_k,u_k,d_k,\epsilon) \equiv |W_yC'(Ax_k+Bu_k+B_dd_k)|-\\|W_y\Big(C(Ax_k+Bu_k+B_dd_k)+C_dd_k\Big)(0.95W_yL_{T_i}+\epsilon)| \leq0.
\label{eq:nl const2}\end{multline}

\begin{figure*}[]
\centering
\includegraphics[scale=0.9]{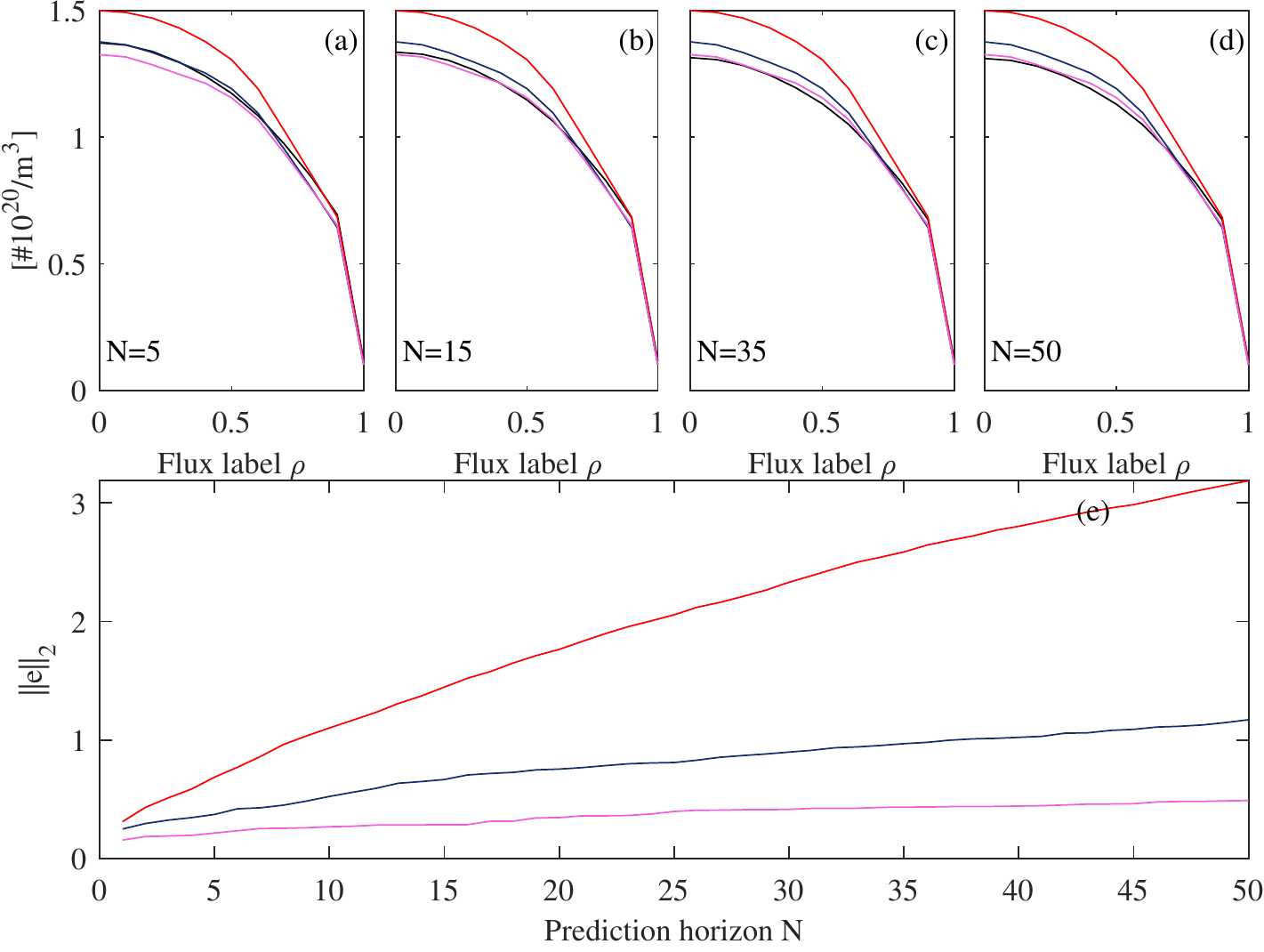}
\caption{Validation of the LDAM and KF in the MPC set-up at different times of the open-loop simulation, i.e., different values of the disturbance estimate $\hat{d}_k$. In red the value of the disturbance after $t=0.25 s$ is used. In blue, the disturbance estimate after $t=1.25 s$ is used and in pink the disturbance estimate at $t=2.5 s$. Frame a-d): The "real" density profile computed by the nonlinear RAPDENS model (black) is compared with the predicted density profile by the LDAM for different values of the prediction horizon N. Frame(e): The prediction error is given as a function of the prediction horizon for different initialization times.}
\label{fig:choice N}
\end{figure*}

\subsection{Cost function matrices}\label{app:cost}
Here we derive the matrices of the numerical implementation of the cost function (\ref{eq:cost function}) as presented in (\ref{eq:optimal control problem}).

First, similar as (\ref{eq:stacked state and input}), we define the stacked steady-state states and inputs as
\begin{equation}
\bar{\bm{X}}_k = \begin{bmatrix}
\bar{x}_{1|k} \\ \bar{x}_{2|k} \\ \vdots \\ \bar{x}_{N-1|k}\\\bar{x}_{N|k}
\end{bmatrix} \; \; \; \; 
\bar{\bm{U}}_k = \begin{bmatrix}
\bar{u}_{0|k} \\ \bar{u}_{1|k} \\ \vdots \\ \bar{u}_{N-2|k}\\\bar{u}_{N-1|k}
\end{bmatrix}.
\label{eq:stacked SS state and input}
\end{equation}
Subsequently, the cost matrices $W_x$, $W_u$, and $W_N$ first introduced in (\ref{eq:cost function}) are used to define the block-diagonal matrices $\mathbb{W}_x$ and $\mathbb{W}_u$ as
\begin{equation}
\mathbb{W}_x = \begin{bmatrix} W_x & 0&\cdots & 0\\0 & W_x & \ddots & \vdots \\\vdots & \vdots & \ddots & \vdots \\0 &0 & \cdots& W_N \end{bmatrix},
\label{eq:Wx}
\end{equation}
\begin{equation}
\mathbb{W}_u = \begin{bmatrix} W_u & 0&\cdots & 0\\0 & W_u & \ddots & \vdots \\\vdots & \vdots & \ddots & \vdots \\0 &0 & \cdots& W_u \end{bmatrix}.
\label{eq:Wu}
\end{equation}
Using (\ref{eq:stacked state and input}), (\ref{eq:stacked SS state and input}), (\ref{eq:Wx}), and (\ref{eq:Wu}) the cost function (\ref{eq:cost function}) can be rewritten as
\begin{equation}
\begin{split} J_k =& x(k)^\top Qx(k) + \bm{X}_k^\top\mathbb{W}_x\bm{X}_k-2\bm{X}_k^\top\mathbb{W}_x\bar{\bm{X}}_k \\&+ \bar{\bm{X}}_k^\top\mathbb{W}_x\bar{\bm{X}}_k+\bm{U}_k^\top\mathbb{W}_u\bm{U}_k-2\bm{U}_k^\top\mathbb{W}_u\bar{\bm{U}}_k \\&+ \bar{\bm{U}}_k^\top\mathbb{W}_u\bar{\bm{U}}_k.\label{eq:compact cost function written out}\end{split}
\end{equation}
Inserting (\ref{eq:Xk}) in (\ref{eq:compact cost function written out}) we obtain:
\begin{align}
J_k = &x(k)^\top Qx(k)+x(k)^\top \Gamma_x^\top\mathbb{W}_x\Gamma_x x(k) \nonumber\\
&+2\bm{U}^\top_k\Gamma^\top\mathbb{W}_x\Gamma_x x(k) + \bm{U}^\top_k \Gamma_u^\top\mathbb{W}_x\Gamma_u\bm{U}_k\nonumber\\
&+2x^\top_k\Gamma_x^\top\mathbb{W}_x\Gamma_dd(k)+2\bm{U}^\top_k\Gamma_u^\top\mathbb{W}_x\Gamma_dd(k) \nonumber\\
&+d(k)^\top\Gamma_d^\top\mathbb{W}_x\Gamma_dd(k)-2x(k)^\top \Gamma_x^\top\mathbb{W}_x\bar{\bm{X}}_k \nonumber\\
&-2\bm{U}_k^\top\Gamma_u^\top\mathbb{W}_x\bar{\bm{X}}_k -2d(k)^\top\Gamma_d^\top\mathbb{W}_x\bar{\bm{X}}_k\nonumber\\
&+\bar{\bm{X}}_k^\top\mathbb{W}_x\bar{\bm{X}}_k+\bm{U}_k^\top\mathbb{W}_u\bm{U}_k-2\bm{U}_k^\top\mathbb{W}_u\bar{\bm{U}}_k \nonumber\\
&+ \bar{\bm{U}}_k^\top\mathbb{W}_u\bar{\bm{U}}_k. \label{eq:reference tracking cost function 2}
\end{align}
Finally, by regrouping the quadratic terms in $\bm{U}_k$, the terms that depend on $\{\bm{U}_k,x(k)\}$, $\{\bm{U}_k,d(k)\}$, and $\{\bm{U}_k,\bar{\bm{X}}_k,\bar{\bm{U}}_k\}$ we can define the matrices $\mathbb{W}_{u,u}$, $\mathbb{W}_{u,x}$, $\mathbb{W}_{u,d}$, and $\mathbb{W}_{u,r}$ from (\ref{eq:optimal control problem}) as
\begin{align*}
 \mathbb{W}_{u,u} &= \mathbb{W}_u+\Gamma_u^\top\mathbb{W}_x\Gamma_u,
 &\mathbb{W}_{u,x} = 2\Gamma_u^\top\mathbb{W}_x\Gamma_x, \\
 \mathbb{W}_{u,d} &= 2\Gamma_u^\top\mathbb{W}_x\Gamma_d,
 &\mathbb{W}_{u,r} = -2\Gamma_u\mathbb{W}_x\bar{\bm{X}}_k-2\mathbb{W}_u\bar{\bm{U}}_k,
\end{align*}

\subsection{Choice of prediction horizon}\label{app:N}

The predictive capacities of the LDAM depend greatly on the estimated disturbance $\hat{d}_k$. The choice of the prediction horizon $N$ is made to trade off control performances and prediction accuracy. In figure \ref{fig:choice N}(a-d) the predicted density profile is shown for different prediction horizons and initialization times. The red lines are obtained with the disturbance estimate $\hat{d}_k$ at time $k=0.25$ s. The blue lines are obtained with the disturbance estimate $\hat{d}_k$ at time $k=1.25$ s. Finally, the pink lines are obtained with the disturbance estimate $\hat{d}_k$ at time $k=2.5$ s.  In (e), the two-norm of the prediction error is shown as a function of the prediction horizon for the three initialization times. 

It can be seen in figure \ref{fig:choice N}(e) that the prediction error increases with the prediction horizon but that is increase is very small when the disturbance estimate has reached the steady-state values (see pink line in \ref{fig:choice N}(e)) and in that case large prediction horizons can be used accurately. In this work, we have chosen to work with N = 20.
\end{appendix}
\end{document}